\documentclass[sigconf]{acmart}

\AtBeginDocument{%
  }

\copyrightyear{2026}
\acmYear{2026}
\setcopyright{cc}
\setcctype{by}
\acmConference[DIS '26]{Designing Interactive Systems Conference}{June 13--17, 2026}{Singapore, Singapore}
\acmBooktitle{Designing Interactive Systems Conference (DIS '26), June 13--17, 2026, Singapore, Singapore}
\acmDOI{10.1145/3800645.3812888}
\acmISBN{979-8-4007-2563-0/2026/06}

\usepackage{multirow}
\usepackage{subcaption}
\usepackage{caption}
\usepackage{xcolor}
\usepackage{framed}
\usepackage{acmart-taps}

\makeatletter
\def\@authorfont{\Large\sffamily}
\def\@affiliationfont{\small\normalfont}
\makeatother

\begin{document}

\title[StreetDesignAI]{StreetDesignAI: {Broadening Designer Perspectives Through Multi-Persona Evaluation of Cycling Infrastructure}}

%\author{Ziyi Wang*}
%\email{zoewang@umd.edu}
%\affiliation{%
%  \institution{University of Maryland, College Park}
%\author{Yilong Dai*}
%\email{ydai17@crimson.ua.edu}
%\affiliation{%
%  \institution{University of Alabama}
%\author{Duanya Lyu}
%\email{lyu.duanya@ufl.edu}
%\affiliation{%
%  \institution{University of Florida}
%\author{Mateo Nader}
%\email{mateo.nader@ufl.edu}
%\affiliation{%
%  \institution{University of Florida}
%\author{Sihan Chen}
%\email{sihanch2@andrew.cmu.edu}
%\affiliation{%
%  \institution{Carnegie Mellon University}
%\author{Wanghao Ye}
%\email{wy891@umd.edu}
%\affiliation{%
%  \institution{University of Maryland, College Park}
%\author{Zijian Ding}
%\email{ding@umd.edu}
%\affiliation{%
%  \institution{University of Maryland, College Park}
%\author{Xiang Yan}
%\email{xiangyan@ufl.edu}
%\affiliation{%
%  \institution{University of Florida}

\author{Ziyi Wang}
\orcid{0009-0006-4821-1255}
\affiliation{
\institution{University of Maryland, College Park}
\city{College Park}
\state{Maryland}
\country{USA}}
\email{zoewang@umd.edu}

\author{Yilong Dai}
\orcid{0009-0005-7459-1768}
\affiliation{
\institution{University of Alabama}
\city{Tuscaloosa}
\state{Alabama}
\country{USA}}
\email{ydai17@crimson.ua.edu}

\author{Duanya Lyu}
\orcid{0000-0002-2468-8888}
\affiliation{
\institution{University of Florida}
\city{Gainesville}
\state{Florida}
\country{USA}}
\email{lyu.duanya@ufl.edu}

\author{Mateo Nader}
\orcid{0009-0009-4042-2682}
\affiliation{\institution{University of Florida}
\city{Gainesville}
\state{Florida}
\country{USA}}
\email{mdnader715@gmail.com}

\author{Sihan Chen}
\orcid{0009-0001-1993-7907}
\affiliation{\institution{Carnegie Mellon University}
\city{Pittsburgh}
\state{Pennsylvania}
\country{USA}}
\email{sihanch2@andrew.cmu.edu}

\author{Wanghao Ye}
\orcid{0009-0002-7064-2335}
\affiliation{
\institution{University of Maryland, College Park}
\city{College Park}
\state{Maryland}
\country{USA}}
\email{wy891@umd.edu}

\author{Zijian Ding}
\orcid{0000-0002-6372-0369}
\affiliation{
\institution{University of Maryland, College Park}
\city{College Park}
\state{Maryland}
\country{USA}}
\email{ding@umd.edu}

\author{Xiang Yan}
\orcid{0000-0002-8619-0065}
\affiliation{\institution{University of Florida}
\city{Gainesville}
\state{Florida}
\country{USA}}
\email{xiangyan@ufl.edu}

\renewcommand{\shortauthors}{Wang, et al.}

\begin{abstract}
Designing cycling infrastructure requires balancing the competing needs of diverse user groups, yet designers often struggle to anticipate how different cyclists experience the same street environment. We investigate how persona-based evaluation can support cycling infrastructure design by making experiential conflicts explicit during the design process. Informed by a formative study with 12 domain experts and crowdsourced bikeability assessments from 427 cyclists, we present StreetDesignAI, an interactive system that enables designers to (1) ground evaluation in real street context through imagery and map data, (2) receive parallel feedback from simulated cyclist personas spanning confident to cautious users, and (3) iteratively modify designs while the system surfaces conflicts across perspectives. A within-subjects study with 26 transportation professionals comparing StreetDesignAI against a general-purpose AI chatbot demonstrates that structured multi-perspective feedback significantly Broaden designers' understanding of various cyclists' perspectives, ability to identify diverse persona needs, and confidence in translating those needs into design decisions. Participants also reported significantly higher overall satisfaction and stronger intention to use the system in professional practice. Qualitative findings further illuminate how explicit conflict surfacing transforms design exploration from single-perspective optimization toward deliberate trade-off reasoning. We discuss implications for AI-assisted tools that scaffold persona-aware design through disagreement as an interaction primitive.
\end{abstract}

%%
%% The code below should be generated by the tool at http://dl.acm.org/ccs.cfm.
%% Please verify and regenerate using the ACM CCS tool before final submission.
%%

%\ccsdesc[500]{Human-centered computing~User studies}
%\ccsdesc[300]{Human-centered computing~Natural language interfaces}


\begin{teaserfigure}
\centering
\includegraphics[width=1\textwidth]{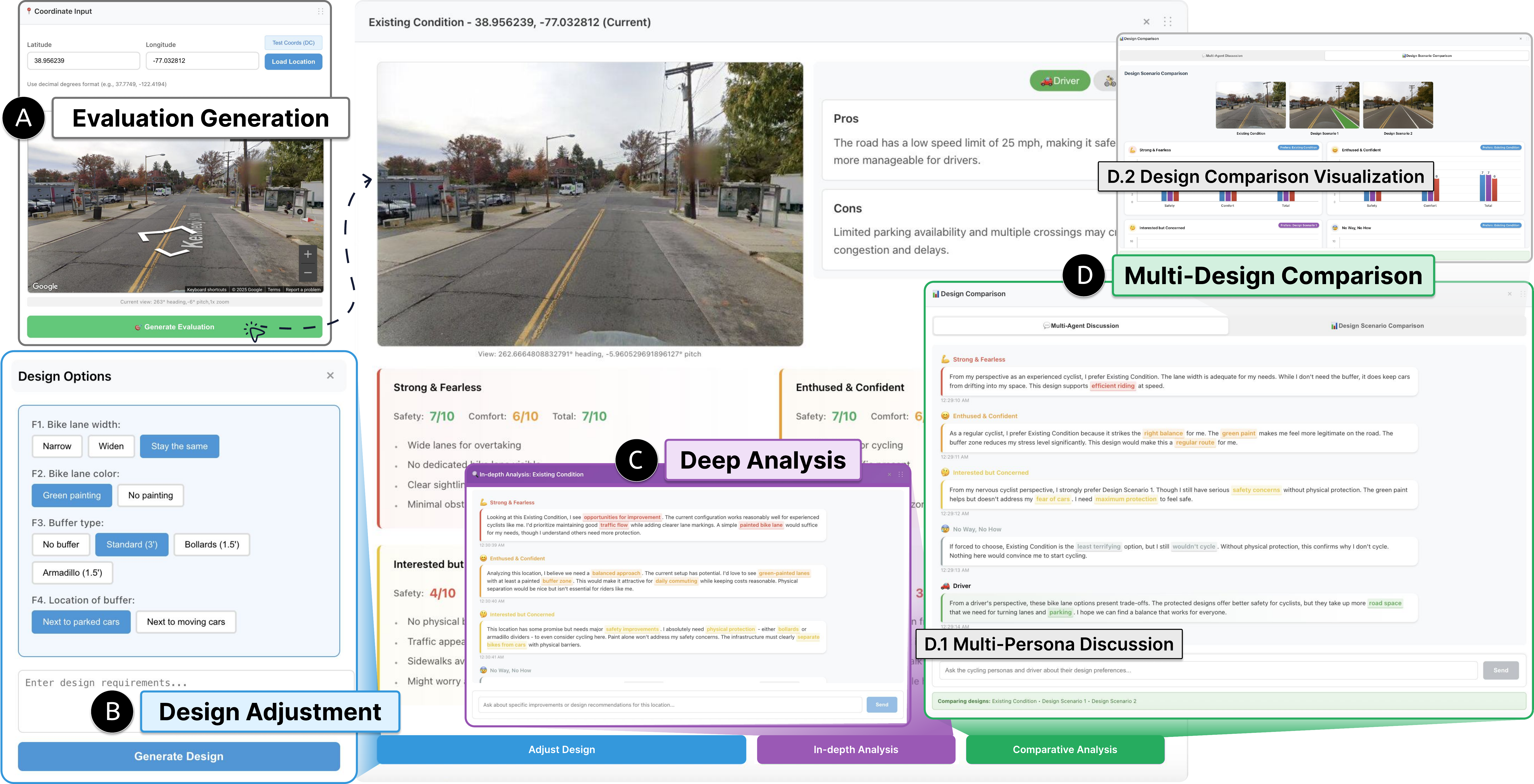}
  \caption{Overview of StreetDesignAI. (A) Users input coordinates to load Street View imagery, which is analyzed using OpenStreetMap data and image recognition. (B) The system generates bikeability evaluations from multiple cyclist personas and allows users to adjust design parameters to create AI-rendered street redesigns. (C) In-depth Analysis provides detailed persona feedback with follow-up questioning capabilities. (D) Comparative Analysis enables multi-design comparison through persona discussions (D.1) and side-by-side score visualizations across safety, comfort, and total metrics (D.2).}
  \Description{Overview of the StreetDesignAI system showing four panels: (A) coordinate input with Street View imagery, (B) multi-persona bikeability evaluations with design parameter controls, (C) in-depth persona feedback with follow-up dialogue, and (D) comparative analysis with persona discussions and side-by-side score visualizations.}
  \label{fig:teaser}
\end{teaserfigure}

\maketitle

\section{Introduction}
\label{sec:intro}

Cycling infrastructure is widely recognized as a cornerstone of sustainable urban transportation~\cite{PUCHER2010S106,pucherWalkingCyclingHealth2010}. When cities build safe and broadly accessible facilities, cycling adoption generally increases and public health benefits follow: protected bike lanes attract substantially more commuters than standard lanes or streets without lanes~\cite{ferenchakLinkLowstressBicycle2025}, and well-designed investments need not worsen traffic congestion. Delineated bicycle lanes may also reduce crash risk and severity for pedestrians and other road users~\cite{YOUNES2024100071}. Yet designing cycling infrastructure that can effectively serve diverse needs remains difficult in practice because designers must negotiate competing demands, limited right-of-way, and heterogeneous public expectations.

A central challenge is that the same street design can be experienced in fundamentally different—and often conflicting—ways by different cyclist populations. Over 60\% of urban residents report willingness to cycle but are deterred by perceived danger in existing conditions~\cite{doi:10.3141/2387-15,FOWLER2017474}, while experienced cyclists may tolerate or even prefer those same facilities for speed and flexibility. Geller's typology highlights four cyclist groups—``Strong and Fearless,'' ``Enthused and Confident,'' ``Interested but Concerned,'' and ``No Way, No How''—whose expectations for safety, comfort, and accessibility diverge sharply~\cite{geller2006four}. The ``Interested but Concerned'' group, representing the largest pool of latent riders, often requires physical protection and clear separation from motor traffic, whereas ``Strong and Fearless'' riders may prioritize directness and maneuverability over protection~\cite{doi:10.3141/2587-11}. These divergent needs create experiential conflicts (e.g., protection vs.\ flexibility) that are difficult to navigate without systematic, context-grounded support.

Existing infrastructure design support methods such as manuals, case databases, expert consultation, and public engagement provide valuable guidance but tend to fall short exactly when conflicts matter most. Design manuals and standards~\cite{nacto2014urban,doi:10.3141/2387-15,aldred2017cycling} encode best practices, but they do not make visible how specific interventions affect diverse users in a particular street context. Public engagement can surface lived experiences, but the process is slow and participation often uneven. Sometimes, by the time sufficient community feedback reaches designers, major constraints are already fixed, leaving little room for meaningful adjustment~\cite{lawlor2023stakeholders}. More recently, AI-based tools have improved visualization and automated technical tasks in urban design, including generative layout exploration~\cite{wang2025generative,quan2019urban, he2025generativeaiurbandesign}, compliance checking~\cite{zhang2023automated}, and perception-informed optimization linking visual features to safety perception~\cite{TAN2025}. However, these tools typically assume a single, homogeneous user perspective; that is, they help designers generate and refine designs, but provide little support for understanding how different user groups experience the same infrastructure or for navigating trade-offs when their needs conflict.

%\textbf{Interaction-level gap.} In early-stage infrastructure design, what designers most need is not more suggestions or a single optimized answer, but a way to treat \emph{disagreement} as a first-class object of design reasoning. Current tools rarely enable designers to systematically \emph{(i)} compare how different user groups experience the same proposal, \emph{(ii)} interrogate why those experiences diverge, and \emph{(iii)} negotiate unavoidable trade-offs under real constraints. In other words, while there are tools that provide guidelines, personas, or visualization, there are few that make \emph{experiential conflict} an \emph{interaction primitive}—something designers can explicitly surface, probe, and work through during iteration.

To address these challenges, we introduce \textit{StreetDesignAI}, an interactive system that enables designers to consider a broader range of cyclist perspectives during cycling infrastructure design. It operationalizes persona-based multi-perspective evaluation as an interaction mechanism to support trade-off reasoning across design options. Designers can use StreetDesignAI to (1) ground evaluation in a specific street context, (2) receive parallel feedback from cyclist personas spanning confident to cautious users, and (3) iteratively modify the design while the system surfaces conflicts and prompts targeted follow-up inquiry. Specifically, designers begin by selecting a street location, after which the system retrieves contextual information from OpenStreetMap and street-level imagery; the system then generates structured evaluations from four cyclist personas, including safety, comfort, and overall bikeability ratings and concise concern summaries. Based on the evaluation results, designers can propose modifications to the street layout via structured parameters (and optional text), receive street-level visualizations of these proposed changes, and obtain updated persona-based evaluations that reveal how each change impacts safety and comfort perceptions across user groups.

We evaluate StreetDesignAI through a within-subjects study of 26 transportation professionals, including students and academic researchers. Participants completed matched street redesign tasks under two conditions, StreetDesignAI and a chatbot baseline, using a counterbalanced within-subjects design. We collected post-task questionnaires and interview data to assess perceived exploration, understanding, confidence, and how conflict surfacing shaped exploration and trade-off reasoning. In particular, we seek to address the following research questions:
\begin{itemize}
\item \textbf{RQ1 - Exploration}: How does persona-based multi-agent evaluation support designers’ exploration of design alternatives and reflection on persona-specific trade-offs?
\item \textbf{RQ2 - Understanding}: How does explicit conflict surfacing across simulated user perspectives influence designers' understanding of diverse user needs compared to general-purpose AI assistance?
\item \textbf{RQ3 - Capability \& Confidence}: How does structured multi-perspective feedback influence designers' perceived capability and confidence in translating diverse user needs into design decisions?
\end{itemize}

The scientific contributions of this work are threefold:
\begin{itemize}
\item We present StreetDesignAI, a grounded, human-informed AI design pipeline that integrates heterogeneous urban context data (e.g., street-level imagery and spatial map data) with real user evaluations to generate persona-aware, situated feedback for more inclusive cycling infrastructure design.

\item We demonstrate that structuring experiential conflicts through persona-based evaluation, comparative analysis, and visualization-driven iteration supports designers’ exploration of design alternatives and deepens their understanding of diverse cyclist personas.

\item Through a mixed-methods, within-subjects study with 26 professionals, we provide empirical evidence that StreetDesignAI improves designers’ perceived capability and confidence in identifying diverse persona needs and translating them into design decisions.

%\textbf{\item we developed a panoramic image-based crowdsourcing system and collected 12,400 persona-conditioned assessments from 427 cyclists and fine-tunes GPT-4.1 to generate persona-specific assessments reflecting lived experience rather than generic or idealized responses.}
\end{itemize}

\section{Related Work}
\label{sec:related}

\subsection{Cycling Infrastructure Design: User Diversity and Experiential Conflict}

Cycling infrastructure design traditionally draws on standardized guidelines, expert judgment, and formal public engagement. Authoritative manuals such as those published by NACTO~\cite{NACTO2011} and AASHTO~\cite{AASHTO1999Bicycle} provide benchmarks for lane dimensions, separation treatments, and intersection design. Transportation research further emphasizes cyclist heterogeneity through frameworks such as Geller’s four cyclist types~\cite{geller2006four} and the Level of Traffic Stress (LTS) model~\cite{Mekuria2012LowStress}, showing how traffic speed, separation, and intersection complexity disproportionately deter less confident cyclists~\cite{furth2013network,montgomery2024lts}. Despite this knowledge, applying user diversity frameworks during design ideation and iteration remains challenging: typologies can remain abstract rather than serving as actionable inputs that guide concrete decisions under constraints~\cite{XIAO2023100949}.

Community engagement methods (e.g., surveys, workshops, hearings) provide another route to incorporate lived experience~\cite{bickerstaff2002transport}, and participatory design frameworks emphasize early-stage exploration of needs through contextual inquiry and iterative prototyping with real users~\cite{beyer1999contextual,brown2008design}. Immersive methods such as 360° video ethnography further enable designers to experience user contexts firsthand, supporting collaborative annotation and iterative sense-making~\cite{meijer2025d360,dai2025using}. However, community participation is frequently a time-intensive process with uneven representation across population groups~\cite{lawlor2023stakeholders}. Notably, these approaches may not consistently support the moment-to-moment design reasoning that an inclusive infrastructure design process requires: surfacing when user experiences conflict, tracing the sources of conflict in a specific context, and negotiating trade-offs before designs harden.

\subsection{AI-Assisted Design: Visualization, Optimization, and the Limits of Single-Perspective Support}

AI-assisted design tools have been adopted to support spatial configuration, optimization, and visualization in infrastructure and urban design~\cite{he2025generativeaiurbandesign,quan2019urban,wang2025generative}. In HCI, human--AI collaboration research frames AI as a partner that complements human judgment~\cite{koch2019may,chen2025genui}, and structured workflows integrating AI into distinct design stages can outperform open-ended conversational assistance~\cite{wu2024framekit,verheijden2023collaborative}. Large language models further enable natural-language interaction and ideation~\cite{liu2025prompts}. Yet when applied to multi-perspective infrastructure design, general-purpose AI tools often provide single-perspective recommendations and may lack mechanisms to reason about conflicting lived experiences and value-laden trade-offs. Designers can ask for advice, but the system does not inherently help them compare stakeholders, interrogate disagreement, or decide what to prioritize.

Visual generation models introduce new opportunities for perception-informed design research. Recent work has explored translating textual descriptions into street-level imagery or editing street scenes through text-based instructions to support perception studies~\cite{deng2024streetscapes,wang2025imagegenerationinfrastructuredesign,TAN2025}. However, perception research requires strict control over variables and visual plausibility of generated content. Applying state-of-the-art image generation models to domain-specific tasks typically demands carefully crafted prompts, parameter tuning, and iterative refinement~\cite{gu2025text2street}, thus practitioners outside the AI field may find it difficult to leverage these models directly for their specific needs.

\subsection{LLM-Based User Simulation: From Multiple Perspectives to Disagreement as an Interaction Primitive}

Recent work explores using LLMs to simulate users for evaluation and feedback, including synthetic usability critique, simulated survey responses, and role-play evaluations~\cite{Lu_2025,argyle2023out,dai2026persona}. Research on Role-Playing Language Agents (RPLAs) provides taxonomies for how LLMs embody diverse perspectives~\cite{chen2024from}, with studies examining how narrative perspective~\cite{li2025eyesee} and persona modality~\cite{kaate2025personas} affect engagement and believability. Multi-agent systems extend this by representing multiple stakeholders~\cite{liu2025personaflow,zhou2024largelanguagemodelparticipatory,quan2026multicolleagues}, and SimTube demonstrates that persona-driven simulation can yield feedback more informative than actual user comments~\cite{hung2025simtube}. However, this literature largely focuses on generating multiple viewpoints rather than making disagreement an operational object. Tools like Synthia show how visual scaffolding helps users act on feedback~\cite{zhang2025synthia}, yet primarily target textual revision. Multi-perspective design can fall short not from lack of diverse opinions, but from lacking mechanisms to surface and negotiate conflicts systematically.

LLM-based simulation also raises credibility concerns: models may produce generic or idealized feedback and misrepresent marginalized users~\cite{argyle2023out}. Fine-tuning, retrieval augmentation, and structured prompting can improve realism~\cite{park2023generative,wang2025careerpooler,park2024generativeagentsimulations1000}. Grounding in real-world data, such as crowdsourced assessments from Project Sidewalk~\cite{saha2019project} with AI-based quality control~\cite{li2024labelaid}, offers one path toward credibility. Multimodal systems like StreetViewAI demonstrate street-level imagery understanding for virtual navigation~\cite{froehlich2025streetviewai}, and recent work explores persona-based urban safety perception~\cite{beneduce2025urbansafetyperceptionlens}. Yet most prior work remains in digital domains, whereas physical infrastructure is embodied and carries long-term consequences. Integrating grounded personas into workflows for early-stage trade-off negotiation remains underexplored.

StreetDesignAI aims to address this gap by supporting explicit negotiation of experiential conflict during iterative infrastructure design. By grounding persona behavior in large-scale cyclist data and coupling evaluation with visualization, StreetDesignAI provides mechanisms for comparing, probing, and resolving disagreement under real-world constraints.

\section{System Design}
\label{sec:system}

StreetDesignAI is an interactive evaluation system that enables designers to consider a broader range of cyclist perspectives during cycling infrastructure design. Our system makes conflicts between stakeholder experiences explicit, supports targeted interrogation of why conflicts arise, and keeps conflicts visible across iteration so designers can negotiate trade-offs.

To achieve this, StreetDesignAI couples three capabilities into one iterative workflow: (1) \textbf{grounded context} (street-view imagery and OpenStreetMap attributes), (2) \textbf{multi-perspective evaluation} through persona-based agents representing heterogeneous user groups, and (3) \textbf{visualization-driven iteration} through parameterized design edits that are rendered at the street level and immediately re-evaluated. This framing distinguishes StreetDesignAI from general-purpose conversational tools (e.g., ChatGPT), which can provide suggestions but do not inherently structure parallel comparison, conflict surfacing, and iterative trade-off negotiation within a single workflow.

We designed StreetDesignAI through a formative study with practitioners and translated those findings into three design goals that guide system features and architecture.

\subsection{Formative Study: Understanding Challenges in Cycling Infrastructure Design}
\label{subsec:formative}

We conducted semi-structured interviews with 12 professionals involved in cycling infrastructure design, including 6 traffic/roadway engineers, 3 transportation major students, 2 transportation planners/project managers, and 1 environmental leaders fellow. All 12 formative study participants later participated in our main user study. Participants reported 1 to 10+ years of experience and involvement in 1 to 5+ cycling-related projects. All interviews were conducted via Zoom and lasted between 17 and 41 minutes ($M = 29.75$, $SD = 7.05$). Our analysis surfaced four recurring challenges that informed system requirements.

\textbf{Challenge 1: Difficulty in perspective-taking beyond personal experience.}
Most participants identified as regular or experienced cyclists and reported difficulty empathizing with less confident populations. As P3 noted: “I bike to work every day and feel comfortable in most conditions... but when I try to imagine how my mother would feel on the same street, I honestly don't know what would concern her.” Participants also described limited familiarity with cyclist typologies (e.g., P7: “I'm not very familiar with the different types of cyclists... I don't really know about the different levels of cyclists.”).

\textbf{Challenge 2: Delayed and limited user feedback in current workflows.}
User feedback typically arrives late (e.g., public hearings or post-implementation feedback), when major constraints are already fixed. P2 emphasized: “By the time we get community feedback, the budget is allocated and the design is mostly fixed... fundamental changes are nearly impossible.” Feedback is also often too coarse to support design trade-offs (P9: “People tell us they want ‘safer bike lanes,’ but that doesn't help us decide between a painted buffer versus physical barriers...”).

\textbf{Challenge 3: Disconnect between technical parameters and lived experience.}
Participants described difficulty translating technical specifications into felt safety/comfort. P11 stated: “I can tell you that a 5-foot bike lane with a 2-foot buffer meets NACTO standards, but I cannot tell you whether a nervous cyclist would feel safe enough to use it.” This disconnect becomes acute under right-of-way constraints (P4: “...choose between widening lanes and improving lane separation... I don't have a good way to understand which matters more for which type of cyclist.”).

\textbf{Challenge 4: Limited access to diverse user perspectives.}
Designers reported that engagement processes tend to over-represent confident cyclists, while hesitant users remain unheard. P6 observed: “Our public meetings tend to attract the same people... We rarely hear from people who don't currently bike but might if conditions improved.” Participants noted that even when they seek broader input, it is slow and hard to make feedback specific to hypothetical designs.

\textbf{Current use of AI tools.}
Eight participants had experimented with ChatGPT or similar tools but reported generic feedback and lack of situated constraint-awareness (P5: “...it gives very generic advice.” P8: “...suggests ideal solutions without acknowledging the constraints we actually face.”). Multiple participants expressed interest in tools that could quickly test how different user types would react to design choices before committing (P12: “What I really need is a way to quickly test how different types of cyclists would react to my design choices, before I've committed to anything.”).

\subsection{Design Goals}
\label{sec:design-goals}

Based on the formative study and prior literature on inclusive design and human--AI collaboration, we derive three design goals:

\textbf{DG1: Facilitate perspective-taking through multi-persona evaluation.}
The system should enable designers to compare heterogeneous user perspectives in parallel and keep differences visible, reducing experience substitution and supporting explicit prioritization decisions.

\textbf{DG2: Support iterative design ideation and concept exploration through real-time visualization and evaluation.}
The system should compress the feedback loop by enabling designers to propose an intervention, visualize it in context, and immediately observe how it shifts conflict patterns across stakeholders.

\textbf{DG3: Bridge technical parameters and lived experience through grounded, narrative feedback.}
The system should translate parameter choices into grounded experiential narratives and visual evidence so designers can reason about why specific elements feel safer or riskier for different users.

\subsection{System Overview}
\label{subsec:overview}

\begin{figure*}[t]
\centering
\includegraphics[width=\textwidth]{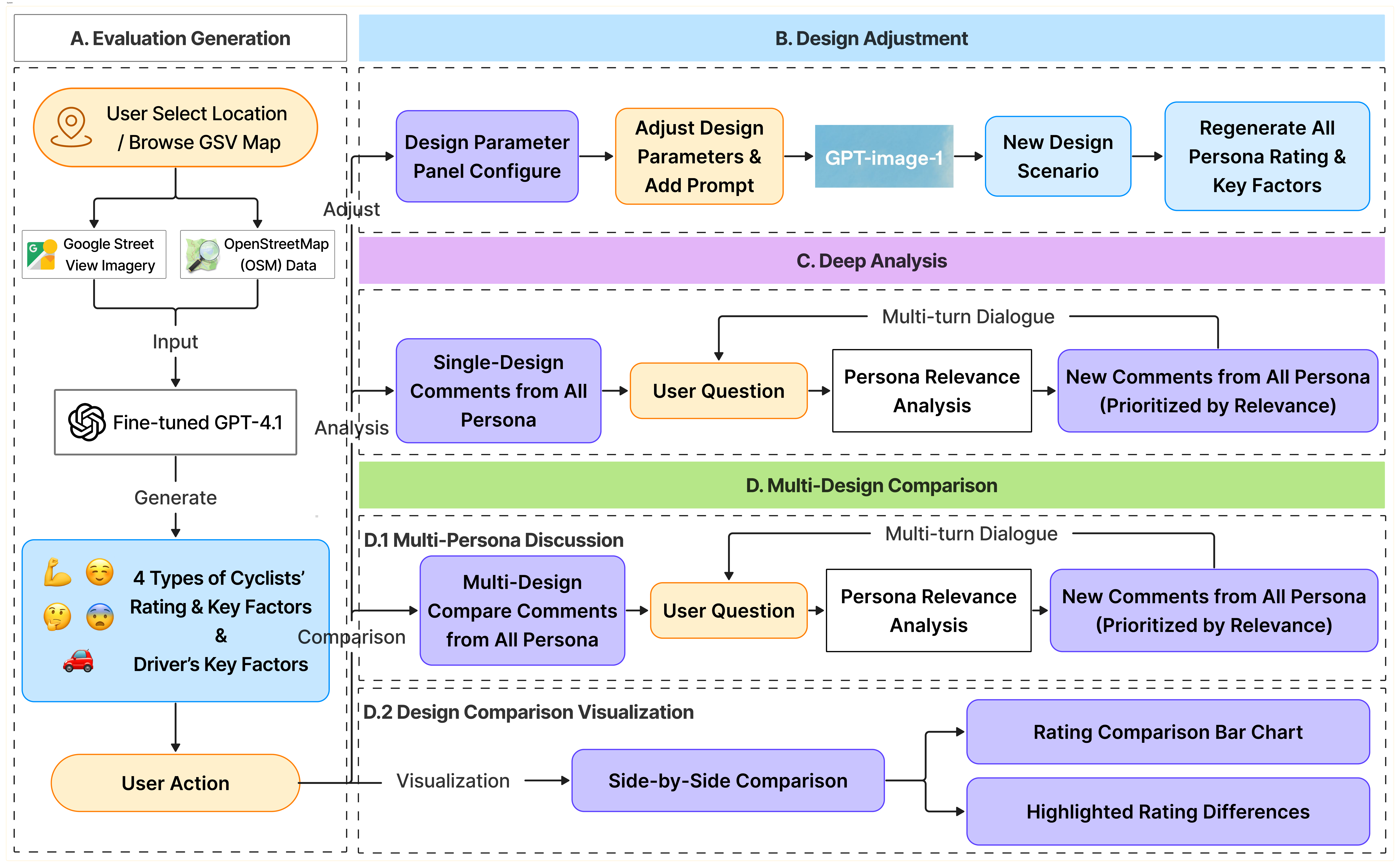}
  \caption{System workflow of StreetDesignAI: The system consists of four main modules: (A) Evaluation Generation collects street-level imagery from Google Street View and road attributes from OpenStreetMap, then uses a fine-tuned GPT-4.1 model to generate safety ratings, comfort scores, and key factors for four cyclist personas and driver's perspective; (B) Design Adjustment allows users to configure design parameters (lane width, color, buffer type) and uses GPT-image-1 to render modified streetscapes, triggering re-evaluation of all personas; (C) Deep Analysis supports multi-turn dialogue for in-depth analysis of design preferences across different personas within a single scenario, with system responses prioritized by relevance; (D) Multi-Design Comparison includes D.1 Multi-Persona Discussion for comparing design preferences across multiple scenarios, and D.2 Design Comparison Visualization module for side-by-side rating comparison with highlighted differences.}
  \Description{System architecture diagram showing four modules: Evaluation Generation retrieves street imagery and map data to produce persona scores; Design Adjustment renders modified streetscapes via GPT-image-1; Deep Analysis enables multi-turn persona dialogue; Multi-Design Comparison provides side-by-side rating visualizations.}
  \label{fig:workflow}
\end{figure*}

StreetDesignAI alternates between \textbf{(1) grounded baseline evaluation} and \textbf{(2) design-and-re-evaluate cycles}. Designers select a real street segment, after which the system retrieves contextual attributes (OpenStreetMap) and street-level imagery (Street View). The system then generates persona-based evaluations representing five stakeholder perspectives:
\begin{itemize}
    \item \textbf{Strong \& Fearless:} experienced daily cyclists prioritizing speed and efficiency
    \item \textbf{Enthused \& Confident:} regular cyclists balancing safety and efficiency
    \item \textbf{Interested but Concerned:} cautious cyclists requiring substantial protection
    \item \textbf{No Way, No How:} non-cyclists deterred by safety concerns, requiring high separation
    \item \textbf{Driver:} operational driving concerns (visibility, turning, lane width, traffic flow)
\end{itemize}

For each iteration, designers specify interventions through structured parameters and optional text. The system generates an edited street-level visualization that reflects those parameters and re-generates evaluations from all personas. Crucially, the system also maintains and surfaces conflict structure across iterations: it highlights where personas diverge, what design elements are driving divergence, and which tensions appear irreducible under constraints, so designers can decide what to prioritize rather than collapsing feedback into a single score.

\subsection{Core Features}
\label{subsec:features}
StreetDesignAI operationalizes disagreement-as-primitive through five integrated features, as shown in Figure \ref{fig:teaser}.

\subsubsection{Evaluation Generation: Parallel baseline assessment from multiple perspectives (DG1, DG2).} Upon street selection via coordinate input, the system retrieves (1) Google Street View imagery and (2) OpenStreetMap attributes (e.g., road class, speed limit, existing cycling infrastructure), as shown in Figure \ref{fig:teaser} (A). Each of the four cyclist personas (Strong \& Fearless, Enthused \& Confident, Interested but Concerned, No Way No How) produces structured feedback including safety scores, comfort scores, and overall ratings (1--10), along with concise observations explaining their assessment. The system also generates driver pros and cons. This parallel evaluation establishes a baseline conflict pattern, making divergent perspectives visible at the outset for subsequent iteration.
\subsubsection{Design Adjustment: Parameterized interventions and AI-rendered visualization (DG2, DG3).}
Designers propose interventions through a Parameter Specification Panel with options for bike lane width (narrow, widen, stay the same), lane color (green painting, no painting), buffer type (standard, bollards, armadillo), and buffer location (next to parked cars, next to moving cars), plus optional free-text requirements, as shown in Figure \ref{fig:teaser} (B). The system uses GPT-Image-1 API to render a modified street view image based on these parameters. After visualization, the system re-runs all persona evaluations on the new design, allowing designers to see how specific parameter changes shift safety and comfort scores across stakeholders.
\subsubsection{Deep Analysis: Interrogating perspectives through conversational follow-up (DG1, DG3).}
Deep Analysis presents each persona's initial observations in a conversation-like format and supports multi-turn dialogue where designers can ask follow-up questions such as ``why do you feel unsafe?'' or ``what changes would improve your rating?'' Personas receive the current image and context, stay in character, reference visible street elements, and provide actionable suggestions, as shown in Figure \ref{fig:teaser} (C). This feature helps designers unpack the mechanisms behind different ratings (e.g., which elements trigger fear for cautious cyclists versus which features drivers find restrictive), moving beyond opaque scores to understand the ``why'' behind evaluations.
\subsubsection{Multi-Persona Discussion: Facilitating cross-persona debate on design preferences. (DG1)}
When comparing multiple design scenarios, the Multi-Agent Discussion tab presents persona responses discussing which design they prefer and why. Each persona articulates their design preference with justification, referencing specific elements (e.g., buffer zones, lane visibility, physical protection), as shown in Figure \ref{fig:teaser} (D.1). Designers can pose questions to the group, and each persona responds in order from high to low according to the relevance of the question. This discussion format surfaces points of agreement, persistent disagreements, and the reasoning behind different preferences, helping designers understand trade-offs that require prioritization rather than simple optimization.
\subsubsection{Design Comparison Visualization: Comparing alternatives through quantitative score profiles (DG1, DG2).}
The Design Scenario Comparison tab enables side-by-side evaluation of the existing condition against generated design scenarios, as shown in Figure \ref{fig:teaser} (D.2). For each persona, the system displays bar charts showing safety, comfort, and total scores across all alternatives, with each design scenario represented by a different color. The visualization also indicates each persona's stated preference at the top. This allows designers to quickly identify which alternatives improve outcomes for specific user groups, where scores diverge across personas, and which design changes produce the largest shifts in perceived safety and comfort---supporting informed decision-making without collapsing diverse perspectives into a single metric.

\subsection{Example Use Scenario}
\label{subsec:scenario}

Zoe is an urban planner redesigning a commercial corridor. She selects a busy arterial with on-street parking and no dedicated cycling facility. StreetDesignAI retrieves street context and produces baseline evaluations: confident cyclists note lack of dedicated space, while cautious personas report they would avoid cycling due to exposure to traffic. The system highlights a baseline conflict pattern: cautious personas demand protection while confident riders and drivers prioritize flexibility and flow.

Zoe proposes a green-painted bike lane with a painted buffer. The visualization shows updated markings and evaluations improve for some personas, but the system indicates that core conflict remains: cautious cyclists still perceive insufficient physical protection. Zoe enters Deep Analysis and asks the Interested but Concerned persona what would make the design usable. The persona requests physical barriers and identifies specific fear points (e.g., proximity to moving vehicles). Zoe adds bollards along the traffic side and observes increased safety for cautious personas, while the driver perspective raises operational concerns. The Multi-Persona Discussion panel marks this as an irreducible trade-off, prompting Zoe to decide what to prioritize under right-of-way constraints. After exploring alternatives, Zoe uses Design Comparison to select an option with a more acceptable conflict profile rather than simply maximizing a single score.

\subsection{Implementation Details}
\label{subsec:implementation}

StreetDesignAI is implemented as a web-based system with a React front end and a model-backed evaluation and visualization pipeline. The system retrieves street-level imagery through the Google Street View API \cite{google_streetview_api} and extracts contextual street attributes through the OpenStreetMap Overpass API \cite{overpass_api_doc} (e.g., road type, signals, existing cycling infrastructure, surrounding features). Designers interact with a custom canvas interface supporting pan/zoom, parameter editing, and multi-view comparison.

For persona-based evaluation and dialogue, the system uses a GPT-4.1 model \cite{openai_gpt4_1_2025} fine-tuned on a crowdsourced bikeability assessment dataset (see Section 3.7 for a detailed description) to reduce generic or idealized responses and improve persona consistency. Each persona agent is defined by a dedicated system prompt specifying identity, priorities, and evaluation focus. Outputs are returned in a structured JSON schema containing scores and observation points to ensure consistency and parsability. All prompts and schemas are included in Appendix~\ref{sec:appendix}.

For visual generation, StreetDesignAI uses GPT-Image-1 \cite{openai2025image} for targeted street view edits. Prompts are structured to identify existing lane regions, apply dimensional specifications tied to selected parameters, and enforce boundary constraints to improve fidelity.

\subsection{Crowdsourced Bikeability Assessment Data Collection}
\label{subsec:data}

To address potential limitations of general-purpose LLMs on domain-specific tasks, we collected bikeability assessment data from real cyclists to ground persona-based evaluations in empirical experience. Specifically, we developed an interactive street rating platform that presents participants with immersive, panoramic Google Street View imagery and asks them to rate perceived safety, comfort, and overall bikeability on 1–10 scales. Open-ended questions are also asked to allow participants to express concerns, preferences, and improvement suggestions. The crowdsourced data are subsequently used to fine-tune GPT-4.1, enabling persona agents to generate evaluations grounded in real cyclist experiences.

\subsubsection{Recruitment And Payment}
We recruited participants through social media platforms to complete anonymous surveys evaluating bicycle lane quality. Each survey took approximately 5 minutes to complete. To incentivize participation, we implemented a lottery-based compensation system: every 100th participant received a \$100 gift card (4 participants total received compensation).  This resulted in an average expected compensation of approximately \$0.93 per participant, or approximately \$11.16 per hour based on the estimated completion time. Participation was voluntary and anonymous, with no personally identifiable information collected. Eventually, a total of 427 cyclists with diverse skill levels, ages, and comfort thresholds completed the survey, yielding approximately 12,400 assessments across varied street conditions.

Three graduate students were recruited from a public R1 doctoral research university to perform data annotation tasks. Annotators were paid \$20 per hour for approximately 8 hours of annotation work each, totaling \$160 per annotator. This compensation rate exceeds local minimum wage standards and is consistent with standard research assistant rates at our institution.

\subsubsection{Data Annotation and Quality Control}
\label{subsubsec:annotation}
Annotators performed two tasks: (1) quality filtering of crowdsourced responses, and (2) classification of open-ended text. For quality filtering, annotators reviewed all submissions and removed responses that were incomplete, nonsensical, or showed clear signs of low effort (e.g., uniform ratings across all items, very short or off-topic text responses). Each unique session was tracked via a session identifier generated when participants opened the survey page. Of 797 total participants who submitted surveys, 576 unique sessions contained actual street assessments; after quality filtering, 427 participants with 12,400 assessments remained from an initial pool of 14,394 ratings. For text classification, annotators independently categorized each open-ended response as expressing concerns about comfort, safety, or neither. Disagreements were resolved through discussion until consensus was reached. These annotations were used to structure fine-tuning data by pairing street context with persona-appropriate evaluation text that reflects category-specific concerns.

\subsubsection{Fine-Tuning}
\label{subsubsec:finetuning}
We fine-tuned GPT-4.1 using the OpenAI fine-tuning API to improve persona-specific evaluation quality. Training examples were constructed by combining street context (location attributes from OpenStreetMap, street-level imagery descriptions) with crowdsourced assessments as target outputs. Each training pair consisted of a system prompt defining the persona identity and evaluation criteria, a user message containing the street context, and an assistant response containing the persona-conditioned evaluation (safety score, comfort score, overall bikeability rating, and qualitative observations drawn from annotated crowdsourced text). Fine-tuning was conducted for 3 epochs with a batch size of 1 and a learning rate multiplier of 2. While we did not perform a formal held-out evaluation comparing fine-tuned outputs against real cyclist assessments of identical streets, the fine-tuned model produced noticeably more situated and persona-consistent responses compared to the base model during iterative development, as assessed qualitatively by the research team. Systematic validation against held-out user assessments is an important direction for future work (see Section~\ref{sec:limitation}).

\subsubsection{Survey Interface}
Figure~\ref{fig:survey1} and Figure~\ref{fig:survey2} show screenshots from our crowdsourcing survey platform. Figure~\ref{fig:survey1} displays the immersive 360-degree Google Street View interface used for bikeability assessment, allowing participants to explore road environments interactively. Figure~\ref{fig:survey2} shows the infrastructure preference assessment interface where participants rate their comfort levels for different cycling facility types.

\section{Study Design}
\label{sec:study}

To evaluate StreetDesignAI's effectiveness in supporting multi-perspective cycling infrastructure design, we conducted a within-subjects comparative study with 26 street designers comparing our system against a conventional AI-assisted design approach and examining how persona-based multi-agent evaluation influences designers' understanding of diverse user needs and design decision-making processes. All sessions were conducted remotely via Zoom, with each session lasting an average of 67.2 minutes (SD = 21.6, range: 34--151 minutes).

\subsection{Participants}
We recruited 26 participants with backgrounds in transportation engineering and planning through email outreach and snowball sampling. Eligibility criteria required participants to have at least one year of experience in road/transportation design or have participated in at least one project involving street design. Participants included practicing professionals such as traffic engineers, transportation planners, and project managers (n=14), academic faculty and researchers (n=4), graduate students (n=6), and undergraduate students (n=2). Participants had varying levels of experience in road/transportation design (M=4.62 years, SD=4.00, range: <1 to >10 years), with 9 participants having more than 5 years of experience. They represented diverse contexts including government agencies, private industry, and academic institutions. Participants ranged in age from 19 to 57 years (M=30.58, SD=9.85), with 18 identifying as male, 7 as female, and 1 as nonbinary. Table~\ref{tab:demographics} provides detailed demographic information.

\begin{figure*}[t]
\centering
\includegraphics[width=\textwidth]{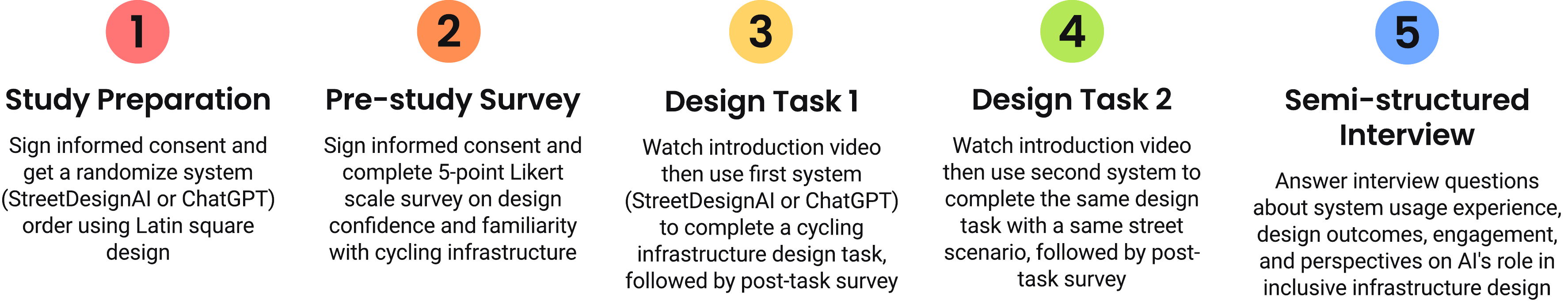}
\caption{Study workflow: participants completed five phases: (1) pre-study survey on design confidence; (2-3) two design tasks using StreetDesignAI and ChatGPT in counterbalanced order, each with post-task surveys; (4) comparative reflection; and (5) semi-structured interview on system experiences and AI's role in multi-perspective design. Total session time averaged 67.2 minutes (SD = 21.6).}
\Description{Study Procedure}
\label{fig:studyprocedure}
\end{figure*}

\begin{table*}[t]
\centering
\caption{Participant demographics and professional backgrounds (N = 26).}
\label{tab:demographics}
\small
\begin{tabular}{|l|c|c|l|c|c|}
\hline
\textbf{ID} & \textbf{Age} & \textbf{Gender} & \textbf{Role} & \textbf{Years} & \textbf{Projects} \\
\hline
P1 & 27 & Female & Transportation Engineering PhD Student & 5-10 & 1-3 \\
P2 & 26 & Male & Transportation Engineering PhD Student & 1-3 & 1-3 \\
P3 & 21 & Male & Transportation Engineering Undergraduate Student & 1-3 & 0 \\
P4 & 21 & Male & Transportation Engineering Undergraduate Student & <1 & 1-2 \\
P5 & 19 & Nonbinary & Environmental Leaders Fellow & <1 & 1-2 \\
P6 & 35 & Male & Traffic Engineer & 5-10 & >5 \\
P7 & 22 & Female & Roadway Engineer 1 & <1 & 1-2 \\
P8 & 27 & Male & Associate Traffic Engineer & 3-5 & 3-5 \\
P9 & 28 & Male & Traffic Engineering Associate & 3-5 & 3-5 \\
P10 & 30 & Male & Transportation Planner & 3-5 & 3-5 \\
P11 & 50 & Male & Sr. Project Manager, County Government & >10 & >5 \\
P12 & 25 & Male & Transportation Engineering PhD Student  & 1-3 & >5 \\
P13 & 25 & Male & Entry Level Transportation Engineer/Urban Designer & 1-3 & >5 \\
P14 & 26 & Male & Transit Planner & 3-5 & 1-2 \\
P15 & 57 & Male & Senior Principal Engineer & >10 & >5 \\
P16 & 27 & Male & Transportation Engineer & <1 & 1-2 \\
P17 & 37 & Female & Transportation Project Manager & >10 & >5 \\
P18 & 29 & Female & Transportation Engineering Faculty & 5-10 & >5 \\
P19 & 53 & Male & Transportation Engineering Professor & >10 & >5 \\
P20 & 29 & Male & Transportation Engineering Research Assistant & 5-10 & 0 \\
P21 & 29 & Female & Transportation Major PhD Student & 1-3 & 3-5 \\
P22 & 23 & Male & Transportation Major Master Student & 1-3 & 0 \\
P23 & 28 & Female & Transportation Engineer & 1-3 & 1-2 \\
P24 & 40 & Male & Transportation Research Faculty & 5-10 & >5 \\
P25 & 37 & Male & Urban Planner & <1 & 1-2 \\
P26 & 24 & Female & Transportation Engineering PhD Student & 1-3 & 1-2 \\
\hline
\end{tabular}
\end{table*}

\subsection{Task Design}

We employed a Latin square balanced within-subjects design to control for potential learning effects. Each participant experienced both conditions:

\textbf{Condition A - StreetDesignAI:} Participants used the full StreetDesignAI system with functions below:
\begin{itemize}
    \item Multi-Persona evaluation from four cyclist types and the driver perspective
    \item Visual generation of design alternatives
    \item Structured parameter controls for infrastructure modifications
    \item Integrated street context from OpenStreetMap
\end{itemize}

\textbf{Condition B - ChatGPT (baseline):} Participants used ChatGPT-4.1 with prompt template that provided street context and requested design recommendations. This baseline represents current practice in AI-assisted road design consultation, where designers might seek general AI guidance without persona-specific perspectives. We selected ChatGPT as our baseline because it is the most widely adopted AI assistant among design professionals (8 of 12 formative study participants had used it), and no existing AI tools specifically support multi-perspective cycling infrastructure evaluation.

For each condition, participants were tasked with improving cycling infrastructure at real street locations. To maintain consistent experimental conditions, all participants used the same two pre-selected street locations, one per condition. The two locations were selected to represent common urban cycling design challenges (e.g., mixed traffic, on-street parking, lack of dedicated cycling facilities) while differing in street context to minimize learning transfer between conditions. The assignment of locations to conditions was counterbalanced across participants using a Latin square design to control for location-specific effects.

\subsection{Study Procedure}
The study procedure followed a structured protocol as illustrated in Figure~\ref{fig:studyprocedure}. Each session began with participants completing a pre-study survey assessing their baseline design confidence and understanding of diverse cyclist needs using adapted items from established scales (e.g., ``I understand how different types of cyclists experience the same infrastructure,'' ``I feel confident designing for users whose needs differ from my own'').

After viewing a brief introduction video about the study objectives (without revealing specific hypotheses), participants proceeded to the first design task. They were instructed to think aloud while exploring design alternatives, documenting their process through screenshots of the three most useful insights or design iterations. This think-aloud protocol provided rich qualitative data about their design reasoning and how they interpreted system feedback.

Following the first condition, participants completed a mid-study survey measuring:
\begin{itemize}
    \item \textbf{Understanding of diverse needs:} Perceived insight into different cyclist perspectives
    \item \textbf{Design confidence:} Self-efficacy in creating solutions that address diverse cyclist needs
    \item \textbf{System usability:} Ease of use and interaction quality
    \item \textbf{Trust in feedback:} Credibility of AI-generated evaluations
\end{itemize}

Participants then repeated the design task using the alternate system, followed by the same assessment battery. The session concluded with a 30-minute semi-structured interview exploring their comparative experiences, perceived value of persona-based feedback, and reflections on AI's role in roadway design practice.

\subsection{Data Collection and Analysis}

We collected multiple forms of data to triangulate findings:

\textbf{Interaction logs:} System telemetry captured all design iterations, parameter modifications, and time spent on different activities. For StreetDesignAI, this included persona feedback viewed, design parameters modified, and iteration patterns. For ChatGPT, we logged conversation turns and design topics discussed.

\textbf{Survey responses:} Likert-scale responses were analyzed using Wilcoxon signed-rank tests to compare paired participant ratings between the two conditions.

\textbf{Interview transcripts:} Following Braun and Clarke's thematic analysis framework~\cite{braunUsingThematicAnalysis2006}, two researchers iteratively coded interview transcripts, identifying patterns in how designers perceived and utilized persona-based feedback. Initial codes were generated inductively, then organized into themes aligned with our research questions. Discrepancies were resolved through discussion until consensus was reached.

\section{Results}
\label{sec:results}

We report results organized by the three research questions. Across all analyses, we combine quantitative survey results with qualitative interview data from 26 participants. For comparisons between StreetDesignAI and ChatGPT, we used Wilcoxon signed-rank tests. For StreetDesignAI-specific features, we report descriptive statistics and one-sample tests against the neutral midpoint. Qualitative findings are used to contextualize and explain observed quantitative patterns.

\begin{table*}[t]
\centering
\caption{Comparison of baseline usability perceptions between ChatGPT and StreetDesignAI. Wilcoxon signed-rank tests compare paired participant ratings under the two systems.}
\label{tab:usabilitycomparison}
\small
\begin{tabular}{lcccc}
\hline
\textbf{Dimension} & \textbf{ChatGPT} & \textbf{StreetDesignAI} & \textbf{W} & \textbf{p-value} \\
 & $(M \pm SD)$ & $(M \pm SD)$ &  &  \\
\hline
Ease of Use & $3.88 \pm 0.86$ & $4.19 \pm 0.85$ & 37.5 & 0.185 \\
Interaction Intuitiveness & $3.58 \pm 0.86$ & $3.88 \pm 0.91$ & 59.5 & 0.216 \\
\hline
\end{tabular}
\begin{flushleft}
\small
 Note: W and p-values represent Wilcoxon signed-rank test results comparing paired participant ratings between ChatGPT and StreetDesignAI.
\end{flushleft}
\end{table*}

\begin{figure*}[t]
 \centering
 \includegraphics[width=1\textwidth]{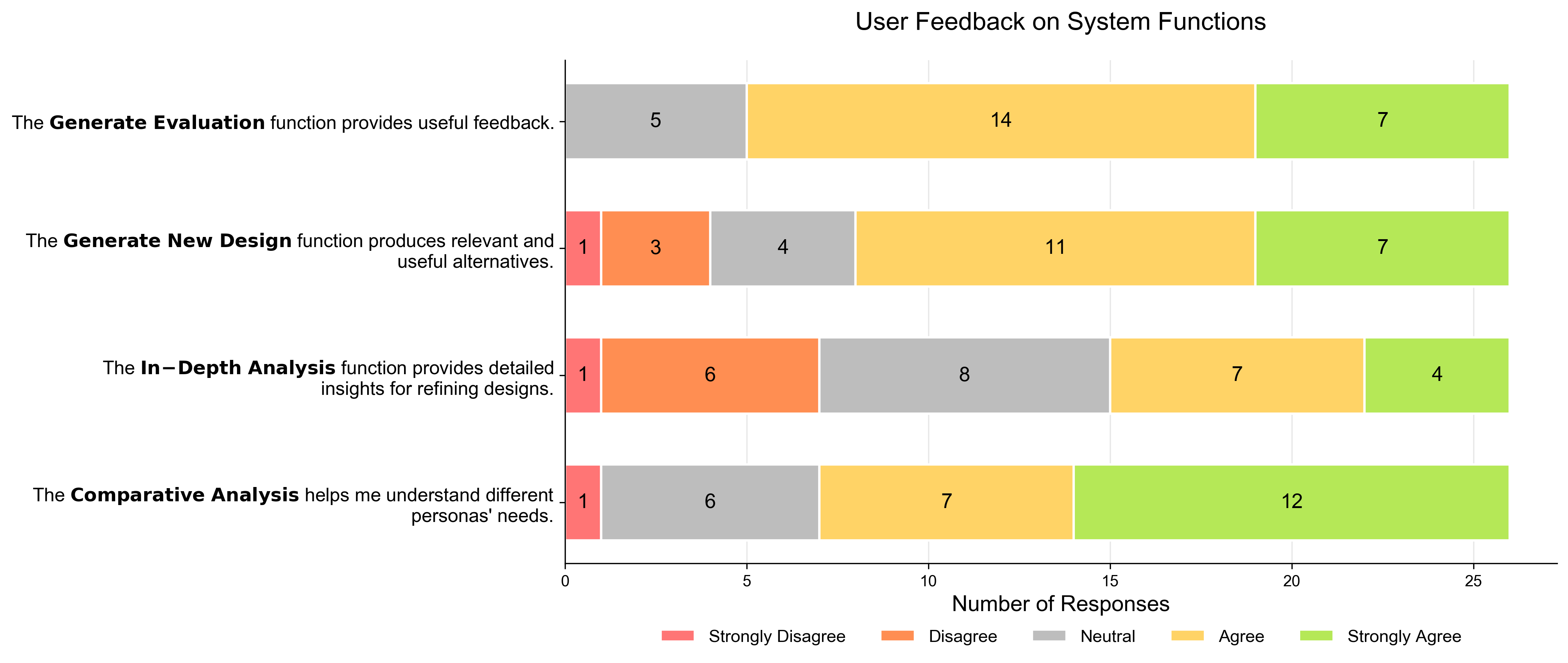}
   \caption{Distribution of participant ratings for four key system functions (N=26). All functions rated above neutral midpoint.}
   \Description{Bar chart showing the distribution of Likert-scale ratings (1-5) for four StreetDesignAI functions: Generate Evaluation, Design Adjustment, In-depth Analysis, and Comparative Analysis. All functions received predominantly positive ratings (4-5).}
   \label{fig:ratingdistribution}
 \end{figure*}

Participants reported comparable levels of baseline usability and interaction intuitiveness for ChatGPT and StreetDesignAI, with no statistically significant differences observed between the two systems (Table \ref{tab:usabilitycomparison}). However, participants rated all StreetDesignAI-specific features significantly above the neutral midpoint (M$=$3), indicating positive perceived support for design exploration, rapid iteration, and trade-off reflection. The Generate Evaluation (M$=$4.31, SD$=$0.79) and Comparative Analysis (M$=$4.46, SD$=$0.71) functions received the highest ratings (Figure \ref{fig:ratingdistribution}).

Participants repeatedly emphasized that StreetDesignAI's structured interface reduced friction in exploration compared to ChatGPT's text-heavy flow. As P12 noted: {``I think that was pretty intuitive. It gave you the image, what's wrong with it, and what these various users would probably think. The scoring system was clear and easy to understand.''}

\subsection{RQ1 - Exploration of Design Alternatives}
\label{subsec:rq1_exploration}

\subsubsection{Rapid Visualization Lowers Barriers to Early-Stage Design Exploration}
\label{sec:visualgeneration}

Rapid visualization enabled designers to explore alternatives early in the workflow. P6 observed that {``within maybe 1-2 minutes, we can visualize changes from existing conditions to proposed improvements. That speed is really valuable for early design exploration.''} Participants described visual generation as both democratizing (P1: ``Not everyone can draw, and even those who can can't produce realistic renderings this fast'') and communicative for stakeholder engagement (P16: ``it helps sell the vision to stakeholders''). While acknowledging occasional inaccuracies in right-of-way constraints (P18) and false positives in infrastructure identification (P25), participants consistently found the capability valuable for concept-level exploration.

Evaluation outputs were perceived as faithful to environmental cues and safety trade-offs. P3 noted that the system {``does a good job understanding what's actually in the environment,''} while P7 confirmed that {``the information it gives makes sense. It correctly identified that there's no bike lane and that the pavement is uneven.''} The speed advantage over manual analysis was salient: P1 stated that {``the AI-generated pros and cons are impressive. It captures the key tradeoffs I would identify manually, but much faster.''}

\subsubsection{Structured Comparison Supports Trade-off Reasoning Across Alternatives}

Comparative analysis strengthened trade-off reflection across alternatives. P10 appreciated that {``the color coding that highlights certain keywords makes it easy to quickly compare feedback across different design scenarios''} and emphasized quantified deltas: {``Seeing the safety score change from 3 to 5 after adding bollards gives clear, quantifiable feedback on design improvements.''} The in-depth analysis feature enabled deeper exploration through follow-up interaction; P18 described it as {``the feature I like most,''} while P2 noted that it {``provides more practical, actionable information. When I asked about separation, it suggested specific buffer types and dimensions.''} Participants also proposed refinements, including more coherent inter-agent dialogue (P25) and better persona-consistent behavior (P18).

\subsubsection{Designers Converge on Visibility and Physical Protection as Core Priorities}

\begin{figure*}[t]
\centering
\includegraphics[width=1\textwidth]{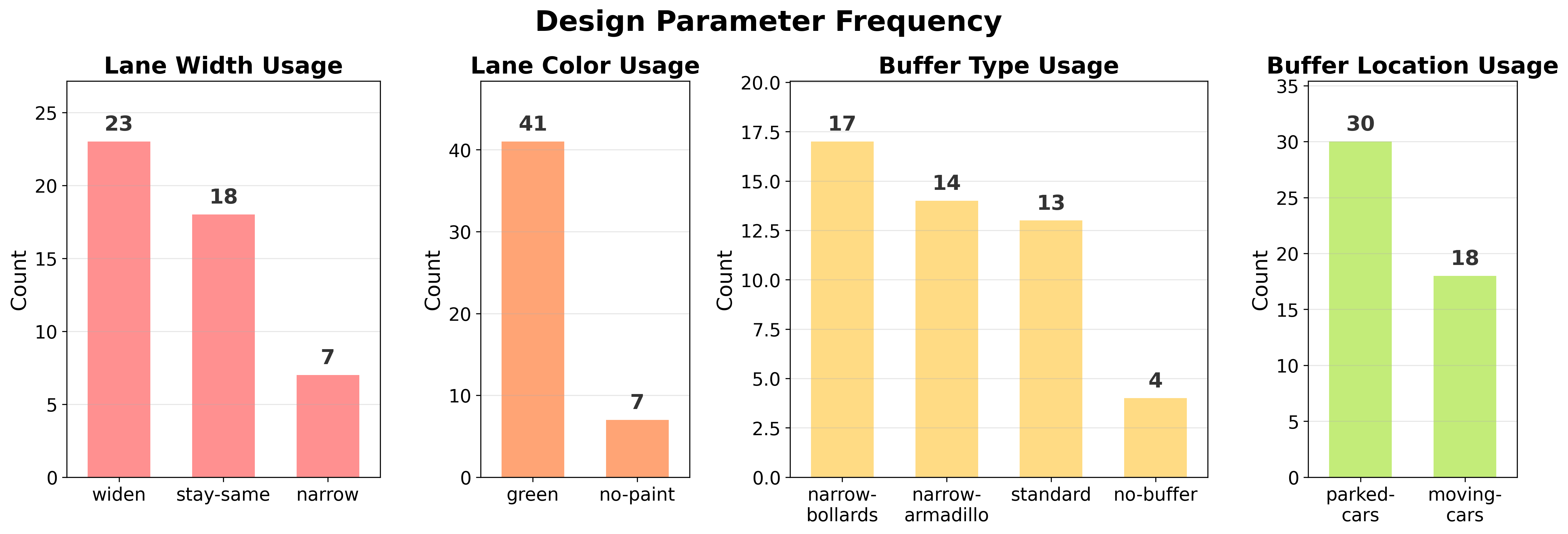}
  \caption{Frequency of design parameter selections across 48 design scenarios: (a) lane width, (b) lane color, (c) buffer type, (d) buffer location.}
  \Description{Four bar charts showing design parameter selection frequencies: (a) lane width preferences with widen most common at 47.9 percent, (b) lane color with green paint dominant at 85.4 percent, (c) buffer type with bollards at 35.4 percent and armadillo at 29.2 percent, (d) buffer location with parked-car side at 62.5 percent.}
  \label{fig:design-parameters}
\end{figure*}

Analysis of 48 design scenarios across 26 sessions revealed consistent patterns in parameter selection (Figure \ref{fig:design-parameters}). Green painted lanes were strongly preferred (85.4\%, 41/48 designs) over unpainted alternatives (14.6\%, 7/48). Physical separation through narrow bollards (35.4\%, 17/48) and armadillo dividers (29.2\%, 14/48) were collectively chosen in 64.6\% of designs, exceeding standard painted buffers (27.1\%, 13/48) and minimal buffering (8.3\%, 4/48). Most participants selected wider lanes (47.9\%, 23/48) over maintaining existing width (37.5\%, 18/48) or narrowing (14.6\%, 7/48). Buffers were more frequently placed adjacent to parked cars (62.5\%, 30/48) than moving traffic (37.5\%, 18/48).

The convergence on green paint (85.4\%) and physical barriers (combined 64.6\%) suggests that visibility and protection emerged as primary design principles. The preference for parked-car-side buffering aligns with participants' explicit concerns about dooring hazards (P8: ``The buffer next to parked cars is essential. I've seen too many door-zone incidents to skimp on that protection''). The diversity of parameter combinations (48 design solutions) demonstrates that StreetDesignAI supported flexible exploration while revealing consistent strategic priorities.

Table \ref{tab:rq1-qual} provides additional qualitative evidence for the themes identified in RQ1.

\begin{table*}[t]
\centering
\caption{RQ1: Qualitative evidence for design exploration themes.}
\label{tab:rq1-qual}
\small
\begin{tabular}{p{0.22\textwidth} p{0.74\textwidth}}
\hline
\textbf{Theme} & \textbf{Participant Quotes} \\
\hline
Visual Generation Speed \& Utility &
P25: ``The image generation is super good. Getting a realistic visualization that quickly is incredibly valuable for concept development.'' \newline
P12: ``If I were in a rush to a meeting and needed to present design concepts, generating these images would be incredibly helpful.'' \newline
P5: ``Image generation has room for improvement, but the current capability is already useful for concept-level discussions.'' \\
\hline
Evaluation Fidelity &
P3: ``The observations about cars are pretty good...how they affect drivers' speed and the hazard of door zones.'' \newline
P15: ``Picked up on subtle details like pavement quality and sight lines that affect cyclist comfort.'' \newline
P25: ``Sometimes it identified bike infrastructure that wasn't there. The false positives could be confusing.'' \\
\hline
Comparison \& Trade-off Reflection &
P3: ``I love the discussions between personas and the comparative analysis. Being able to see how each group responds to changes is very useful.'' \newline
P2: ``Having a basic idea in my head, this gives me a more intuitive comparison than I could do mentally.'' \newline
P16: ``Helped me understand which design elements had the biggest impact on different user groups.'' \\
\hline
Iterative Exploration &
P1: ``You can follow up with questions and dig deeper into specific concerns. The interactivity adds value.'' \newline
P18: ``The `strong and fearless' persona could be less focused on physical barriers since they're supposed to be comfortable with less protection.'' \\
\hline
Structured Interface &
P5: ``The interface on the website is much more intuitive than ChatGPT. It allows you to view each scenario more clearly as its own section instead of scrolling through text.'' \newline
P5: ``Instead of filling out a text box, selecting buttons is much more efficient. The structured approach helps organize my thinking.'' \\
\hline
\end{tabular}
\end{table*}

\subsection{RQ2 - Understanding of Diverse User Needs}
\label{subsec:rq2_understanding}

\subsubsection{Multi-Persona Feedback Broadens Understanding Beyond Personal Experience}

\begin{table*}[t]
\centering
\caption{Comparison of perceived understanding and intention-to-use outcomes between ChatGPT and StreetDesignAI. Wilcoxon signed-rank tests compare paired participant ratings under the two systems.}
\label{tab:understanding-comparison}
\small
\begin{tabular}{lcccc}
\hline
\textbf{Dimension} & \textbf{ChatGPT} & \textbf{StreetDesignAI} & \textbf{W} & \textbf{p-value} \\
 & $(M \pm SD)$ & $(M \pm SD)$ &  &  \\
\hline
Understanding new perspectives
& $3.46 \pm 0.99$
& $4.04 \pm 1.00$
& 43.5
& 0.031* \\

Authenticity of persona feedback
& $3.31 \pm 1.01$
& $3.77 \pm 0.71$
& 51.5
& 0.070 \\

Overall Satisfaction
& $3.38 \pm 0.85$
& $3.85 \pm 0.83$
& 24.0
& 0.034* \\

Intention to use in professional work
& $2.88 \pm 1.34$
& $3.62 \pm 1.13$
& 46.0
& 0.025* \\
\hline
\end{tabular}
\begin{flushleft}
\small
 Note: * $p < 0.05$. W and p-values represent Wilcoxon signed-rank test results comparing paired participant ratings between ChatGPT and StreetDesignAI.
\end{flushleft}
\end{table*}

StreetDesignAI showed significantly higher ratings than ChatGPT in helping participants understand perspectives they had not previously considered in street design (W$=$43.5, $p=0.031$), with mean ratings increasing from 3.46 to 4.04 (Table \ref{tab:understanding-comparison}). StreetDesignAI also led to significantly higher overall satisfaction (3.38 to 3.85, W$=$24.0, $p=0.034$) and stronger intention to use the system in professional work (2.88 to 3.62, W$=$46.0, $p=0.025$). While participants rated persona feedback from StreetDesignAI as more authentic than ChatGPT (3.31 to 3.77), this difference did not reach statistical significance ($p=0.070$).

\subsubsection{Persistent Score Stratification Reveals Limits of Standard Interventions}

\begin{figure*}[t]
\centering
\includegraphics[width=1\textwidth]{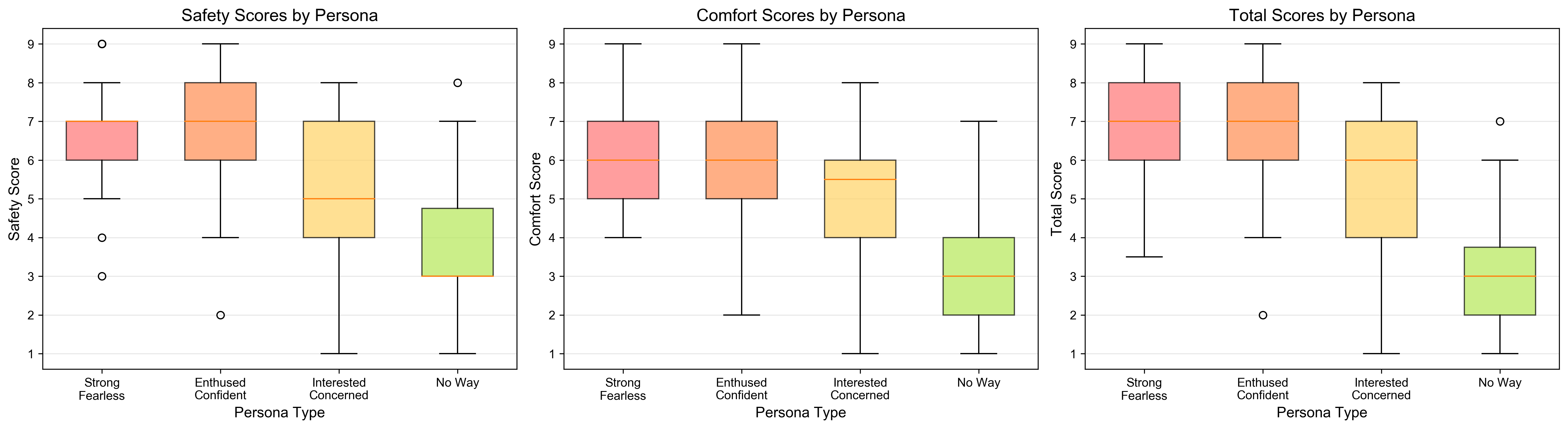}
  \caption{Distribution of safety, comfort, and overall suitability scores across four cyclist personas (N=78 evaluations from 26 sessions).}
  \Description{Box plots showing score distributions for four cyclist personas across safety, comfort, and overall suitability. Strong and Fearless and Enthused and Confident cluster around 7, Interested but Concerned around 5-6 with high variability, and No Way No How consistently below 4.}
  \label{fig:persona-scores}
\end{figure*}

Analysis of 78 evaluations (30 existing conditions + 48 design scenarios) revealed systematic divergence in how different personas rated infrastructure (Figure \ref{fig:persona-scores}). Strong \& Fearless (M$=$6.90, SD$=$1.08) and Enthused \& Confident (M$=$6.91, SD$=$1.22) cyclists rated scenarios similarly, with median overall scores around 7 and relatively compact distributions. In contrast, Interested but Concerned cyclists showed substantially lower scores (M$=$5.58, SD$=$1.76) with high variability, indicating that this group's experience is highly sensitive to design choices. No Way No How cyclists rated nearly all scenarios below 4 (M$=$3.02, SD$=$1.23), with minimal variation across scenarios. The 3.9-point gap between confident cyclists (M$\approx$6.9) and the most cautious group (M$=$3.02) quantifies the magnitude of experiential conflict that designers must navigate.

\begin{table*}[t]
\centering
\caption{Pre--post changes in designers' perceived capability and confidence under ChatGPT and StreetDesignAI conditions. Significance tests compare pre-study baseline measurements with post-condition assessments using Wilcoxon signed-rank tests.}
\label{tab:rq3-capability-comparison}
\resizebox{\textwidth}{!}{%
\begin{tabular}{lcccccccc}
\hline
\multirow{2}{*}{\textbf{Dimension}} & \multirow{2}{*}{\textbf{Pre-study}} & \multicolumn{3}{c}{\textbf{ChatGPT}} & \multicolumn{3}{c}{\textbf{StreetDesignAI}} \\
\cline{3-5} \cline{6-8}
 &  & \textbf{Post} & \textbf{W} & \textbf{p-value} & \textbf{Post} & \textbf{W} & \textbf{p-value} \\
 & $(M \pm SD)$ & $(M \pm SD)$ &  &  & $(M \pm SD)$ &  &  \\
\hline
Persona Need Identification Capability
& $3.27 \pm 1.04$
& $3.73 \pm 0.78$
& 48.0
& 0.044*
& $4.12 \pm 0.91$
& 18.5
& 0.0028** \\

Design Translation Confidence
& $3.27 \pm 1.22$
& $3.81 \pm 0.90$
& 50.0
& 0.063
& $4.19 \pm 0.75$
& 25.5
& 0.0045** \\
\hline
\end{tabular}%
}
\begin{flushleft}
\small
 Note: * $p < 0.05$, ** $p < 0.01$. W and p-values represent Wilcoxon signed-rank test results comparing pre-study baseline scores with post-condition assessments.
\end{flushleft}
\end{table*}

\begin{figure*}[t]
\centering
\includegraphics[width=0.9\textwidth]{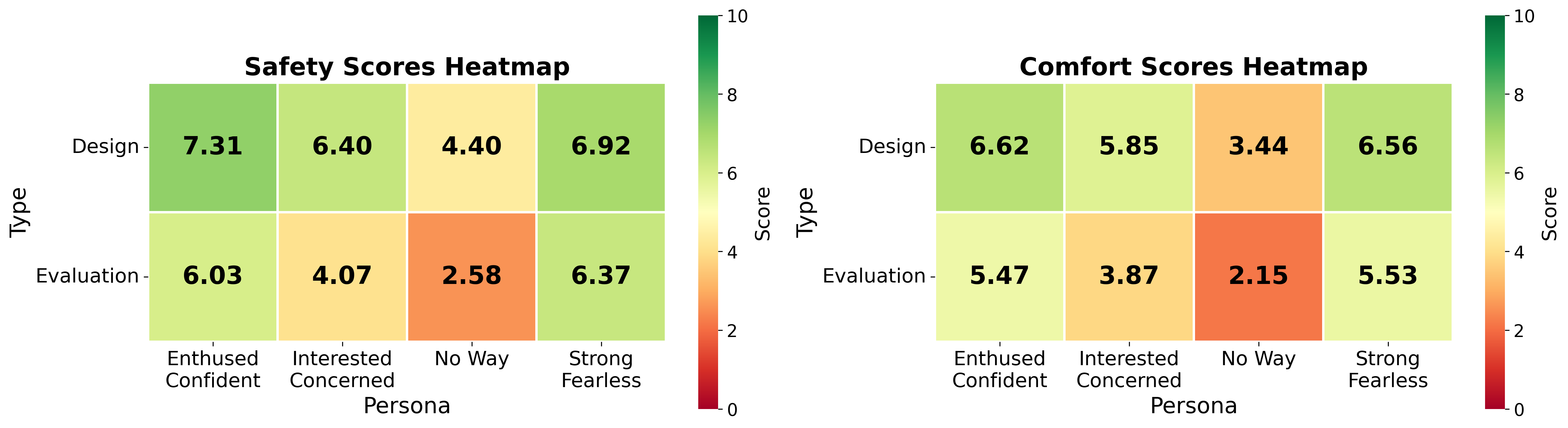}
  \caption{Mean safety and comfort scores by persona and scenario type (evaluation vs. design). Color intensity indicates score magnitude (red = lower, green = higher).}
  \Description{Heatmap comparing mean safety and comfort scores across four personas for evaluation versus design scenarios. Confident personas show green (high scores) while No Way No How shows red (low scores) across both scenario types, with modest improvements in design scenarios.}
  \label{fig:persona-heatmap}
\end{figure*}

Comparing existing condition evaluations against design scenario proposals revealed differential responsiveness to interventions (Figure \ref{fig:persona-heatmap}). Design scenarios consistently received higher scores than evaluations across most personas. However, improvement magnitude varied substantially: Interested but Concerned showed measurable gains in comfort (5.13 in evaluations $\rightarrow$ 5.30 in designs, +0.17), while No Way No How showed no improvement and slight declines (safety: 3.74 $\rightarrow$ 3.67; comfort: 2.97 $\rightarrow$ 2.91). The persistent stratification---confident cyclists rating 3+ points higher than cautious cyclists even in improved designs---underscores the challenge of achieving universal satisfaction through standard protected lane treatments. As P15 observed: {``Even with physical barriers, some people just won't feel safe sharing road space with cars. They need complete separation, like a sidepath or protected cycle track.''}

Table \ref{tab:rq2-qual} provides additional qualitative evidence for the perspective-taking and authenticity themes identified in RQ2.

\begin{table*}[t]
\centering
\caption{RQ2: Qualitative evidence for perspective-taking and perceived authenticity.}
\label{tab:rq2-qual}
\small
\begin{tabular}{p{0.22\textwidth} p{0.74\textwidth}}
\hline
\textbf{Theme} & \textbf{Participant Quotes} \\
\hline
Surfacing Overlooked Perspectives &
P4: ``Perspectives I hadn't considered. For example, it was the only one to acknowledge the available sidewalk as an alternative for cautious cyclists.'' \newline
P3: ``The personas are really useful because I don't usually think about different cyclist comfort levels when doing road design.'' \newline
P1: ``Can remind you that a certain design might actually make some user groups feel worse, even when overall scores improve. That's a critical insight.'' \newline
P5: ``It made me think about weather in ways I hadn't really considered before, like how rain or snow can affect lane markings and surface traction.'' \\
\hline
Contextual \& Situational Understanding &
P15: ``People can be in multiple categories depending on context. I was a very different cyclist when riding with my younger children. The tool captures that nuance.'' \newline
P5: ``The `no way, no how' persona is like a mother's perspective of a child taking this route to school... That's a real user group we need to design for.'' \newline
P4: ``People who are worried about cycling risk won't be convinced by minor improvements, they need substantial infrastructure changes.'' \newline
P7: ``As someone who's comfortable on most roads, I tend to forget how intimidating they can be for less confident cyclists.'' \\
\hline
Perceived Authenticity &
P5: ``The personas stay pretty true to their perspective. The `no way, no how' and `interested but concerned' especially feel authentic.'' / ``It mimics a real person pretty accurately.'' \newline
P2: ``The responses are fairly predictable given each persona's profile, which actually increases credibility.'' \newline
P18: ``You can tell it's AI-generated if you read carefully, but the substance of the feedback is valid and useful regardless of the source.'' \\
\hline
Qualitative over Numeric Value &
P25: ``The qualitative comparison is more insightful than just rating safety 7 versus 6. Understanding why different users feel differently is more valuable than the numbers alone.'' \newline
P5: ``I'd rate it nine out of ten for gaining perspective on how people might perceive a street redesign from both driver and cyclist viewpoints.'' \\
\hline
\end{tabular}
\end{table*}

\subsection{RQ3 - Capability and Confidence in Multi-Perspective Design}
\label{subsec:rq3_capability}

\subsubsection{Structured Persona Feedback Improves Designer Capability and Confidence}

Both ChatGPT and StreetDesignAI improved participants' ability to identify and articulate the needs of different cyclist personas, with stronger improvement observed under StreetDesignAI (Table \ref{tab:rq3-capability-comparison}). Persona need identification increased from baseline (M$=$3.27, SD$=$1.04) to post-ChatGPT (M$=$3.73, SD$=$0.78, W$=$48.0, $p=0.044$) and further to post-StreetDesignAI (M$=$4.12, SD$=$0.91, W$=$18.5, $p=0.0028$). Only StreetDesignAI produced a significant increase in confidence translating persona needs into design decisions (baseline M$=$3.27, SD$=$1.22 $\rightarrow$ post M$=$4.19, SD$=$0.75, W$=$25.5, $p=0.0045$), compared to ChatGPT's non-significant change (post M$=$3.81, SD$=$0.90, $p=0.063$). These findings suggest that while general-purpose AI tools may support initial awareness of diverse user needs, StreetDesignAI more effectively supports the translation of those needs into actionable design decisions. We note that this increased confidence reflects participants' self-reported perceptions rather than an objective measure of design quality; whether the resulting designs better serve diverse cyclist populations requires further validation (see Section~\ref{sec:limitation}).

\subsubsection{Granular Feedback Informs Persona-sensitive Design Decisions}

\begin{figure*}[t]
\centering
\includegraphics[width=1\textwidth]{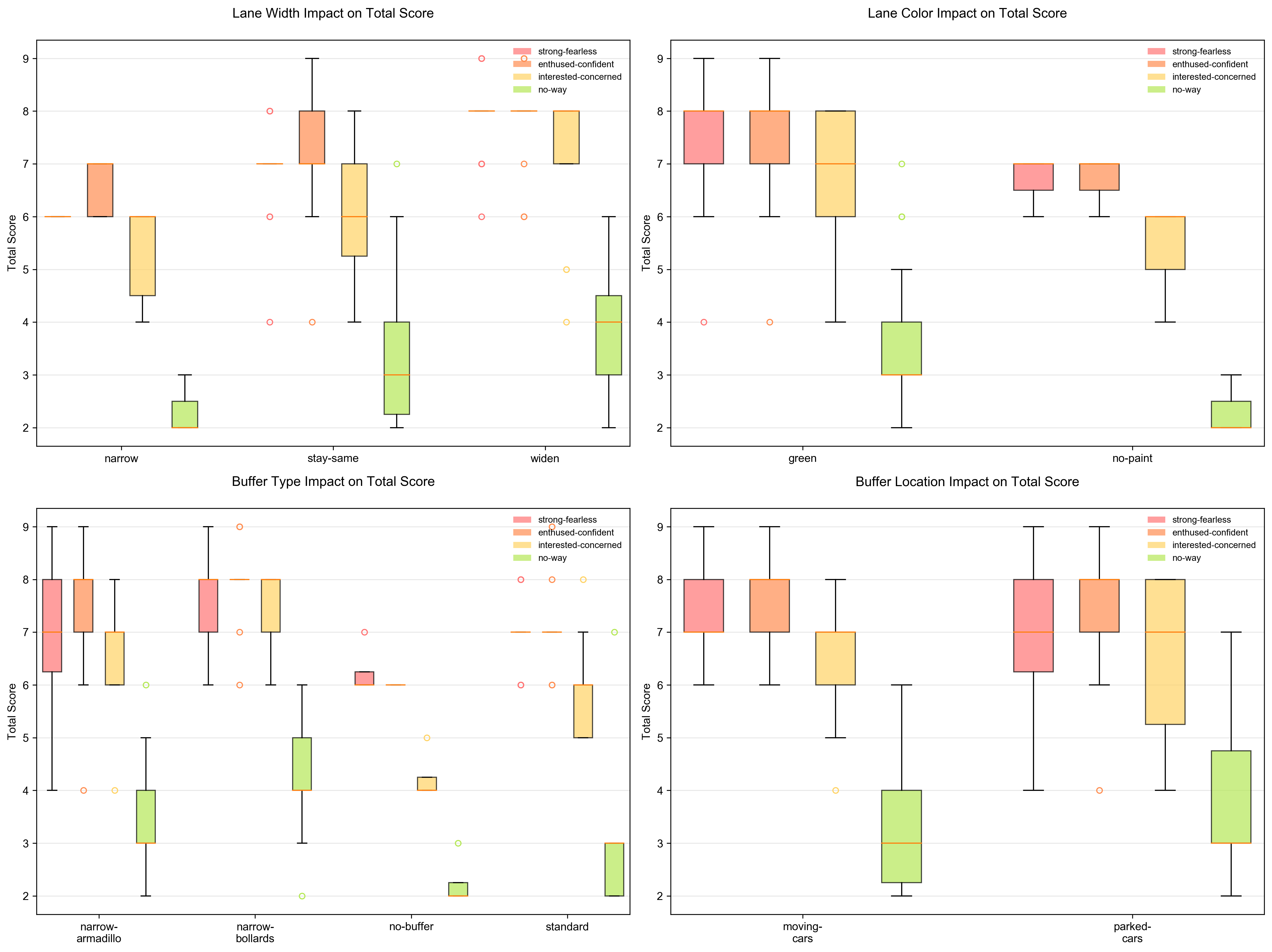}
  \caption{Distribution of overall suitability scores by design parameter choice and persona: (a) lane width, (b) lane color, (c) buffer type, (d) buffer location.}
  \Description{Four grouped box plot panels showing how design parameters affect persona scores. Interested but Concerned shows the steepest gains from widened lanes, green paint, and physical barriers. No Way No How remains below 4 regardless of parameter choice.}
  \label{fig:parameter-impact}
\end{figure*}

Examining how specific design parameters influenced persona scores revealed differential sensitivities that inform targeted interventions (Figure \ref{fig:parameter-impact}). Lane width showed progressive improvements across all personas, with Interested but Concerned exhibiting the steepest increase (median $\approx$ 6.5 for widened lanes vs. 4.5 for narrow lanes, approximately +2.0 points). Green paint substantially benefited Interested but Concerned cyclists (median $\approx$ 6.0 vs. 4.0 without paint, +2.0 points) while having minimal effect on Strong \& Fearless, indicating that visibility enhancements disproportionately serve cautious users. Physical barriers (bollards and armadillos) elevated Interested but Concerned scores (median $\approx$ 7.0) compared to standard buffers (median $\approx$ 5.0) or no buffer (median $\approx$ 3.0), representing approximately +2.0 points improvement. However, No Way No How remained below 4.0 regardless of buffer type, suggesting that standard protected lanes may be insufficient for the most risk-averse population. Buffer location showed modest effects, with parked-car-side placement yielding slightly higher scores for Interested but Concerned, consistent with prioritizing dooring risk mitigation.

These granular insights illustrate how StreetDesignAI can support evidence-based parameter selection: designers can identify which interventions serve which populations, avoiding one-size-fits-all solutions. The variation in sensitivity across personas clarifies which design elements address which concerns: green paint enhances legitimacy and visibility (key for Interested but Concerned), physical barriers provide psychological reassurance (essential for cautious riders), and width supports both maneuverability (valued by Strong \& Fearless) and personal space (critical for nervous cyclists).

Table \ref{tab:rq3-qual} provides additional qualitative evidence for the capability and confidence themes identified in RQ3.

\begin{table*}[t]
\centering
\caption{RQ3: Qualitative evidence for capability gains and confidence mechanisms.}
\label{tab:rq3-qual}
\small
\begin{tabular}{p{0.22\textwidth} p{0.74\textwidth}}
\hline
\textbf{Theme} & \textbf{Participant Quotes} \\
\hline
Educational Value (Novices) &
P13: ``As a beginner planner in the multimodal space, I found the structured personas very educational. It taught me how to think about different user types.'' \newline
P18: ``For planners specializing in land use or housing, this would be even more helpful.'' \\
\hline
Corrective Value (Experts) &
P15: ``Even as an experienced professional, I found value in the systematic approach. It's easy to develop blind spots, and this tool helps identify them.'' \newline
P1: ``The persona framework considers diverse cyclists more comprehensively than I would on my own. With ChatGPT, I'd need to explicitly prompt for each perspective.'' \newline
P6: ``The tool bridges the gap between technical knowledge and user experience. It translates engineering concepts into human impacts.'' \\
\hline
Visual Feedback \& Optimization &
P12: ``If I needed to decide whether to put the bike lane inside or outside parked cars, I could generate these images and see what it would look like. That visual feedback builds confidence.'' \newline
P2: ``Now I can ensure the first three user types all score above 7, then find the lowest cost solution that achieves that. It gives me a clear optimization target.'' \newline
P15: ``The tool gave me confidence that I'm not missing major concerns. Having systematic feedback from multiple perspectives is reassuring.'' \\
\hline
Stakeholder Justification &
P16: ``More prepared to defend my design choices now. I can anticipate objections and explain how I've considered different user needs.'' \newline
P3: ``This provides a good baseline for what options to show people and what questions to bring up in public engagement.'' \\
\hline
\end{tabular}
\end{table*}

\section{Discussion}
\label{sec:discussion}

In this section, we discuss the implications of our findings for AI-assisted multi-perspective design, situated in past research on human-AI collaboration and infrastructure planning. We also offer design recommendations for future systems utilizing persona-based evaluation.

\subsection{Conflict Surfacing as a Design Primitive for Multi-Perspective Infrastructure Design}

A central finding of this work is that making experiential conflicts explicit reshapes how designers explore and reason about alternatives. Beyond past studies focusing on general-purpose AI design assistance~\cite{he2025generativeaiurbandesign}, we found evidence that StreetDesignAI's multi-persona design can enhance multi-perspective design reasoning by directly presenting designers with preference divergences among different user groups. This design choice aligns with calls in HCI research for AI systems that support rather than supplant human judgment~\cite{chen2025genui, koch2019may, bai2026haid}. P1 observed that the persona system ``can remind you that a certain design might actually make some user groups feel worse, even when overall scores improve,'' articulating the value that conflict visibility may help prevent false consensus and prompt designers to take responsibility for their prioritization decisions. By further examining participants' design exploration behaviors (Section \ref{subsec:rq1_exploration}), we identified that explicitly presenting preference divergences provides designers with a means to recognize trade-offs requiring deliberate choices rather than technical optimization. This design promotes user agency while supporting the ideation process, drawing from the experiential knowledge of different cyclist populations, meanwhile reducing perspective substitution and addressing concerns about designing only for users similar to oneself.\enlargethispage{16pt}

\subsection{Cultivating Trade-off Reasoning through Multi-Perspective Evaluation}

Our findings indicate that StreetDesignAI's persona-based design not only supports perspective-taking but also triggers higher-order reasoning processes crucial for cycling infrastructure design. In particular, we observed participants repeatedly engaging in explicit prioritization activities when persona evaluations diverged. StreetDesignAI's design of presenting parallel persona feedback prompts participants to think about whose needs to prioritize under various constraints, thereby stimulating trade-off reasoning.

For instance, by viewing divergent persona evaluations, participants reformulated their design goals in more precise terms (e.g., shifting from ``improve cycling conditions'' to ``expand access for cautious cyclists while maintaining efficiency for confident riders''). Comparison instances (e.g., where participants identified that bollards improved ``Interested but Concerned'' scores while ``No Way No How'' scores remained unchanged) were associated with explicit decisions about which populations to invest in serving.

Additionally, the system's comparative analysis structure and iterative parameter adjustment mechanism promoted designer-led reflection through inference activities based on score changes across design iterations. Our findings suggest that this conflict-surfacing interaction design contributes to designers' deliberation and reasoning process by promoting different levels of trade-off activities. These observations highlight the potential application of persona-based evaluation for fostering deliberate prioritization in infrastructure design and planning education~\cite{beyer1999contextual, brown2008design}.

\subsection{From Understanding to Translation: Bridging the Confidence Gap}

Our quantitative results reveal an important distinction between understanding diverse user needs and confidently translating that understanding into design decisions. While both ChatGPT and StreetDesignAI improved participants' ability to identify persona needs, only StreetDesignAI produced significant gains in translation confidence ($p=0.0045$ vs. $p=0.063$). This gap suggests that awareness of user diversity is necessary but insufficient for effective multi-perspective design practice---designers still need scaffolding to act on that awareness.

StreetDesignAI's parameter-level feedback appears to bridge this gap by linking specific design choices to persona-specific outcomes. When P6 observed that adding bollards increased ``Interested but Concerned'' scores from 2 to 5 while ``No Way No How'' scores remained unchanged, they gained actionable insight into which interventions serve which populations. StreetDesignAI can help translate abstract understanding into targeted decision-making, helping designers more systematically address diverse user needs.

\subsection{Navigating Potential Biases and Limitations in Persona-Based Evaluation}

Cognitive biases have been common concerns in LLM-based user simulation systems~\cite{argyle2023out}. Although our findings demonstrated the utility of the system in providing multi-perspective design support, we noted potential biases when designers interacted with persona evaluations. In StreetDesignAI, confirmation bias may arise from both designers' persona attention patterns and the LLM's inherent limitations. Participants who identified as experienced cyclists tended to engage more deeply with the ``Strong \& Fearless'' persona feedback, potentially reinforcing their existing design intuitions. Additionally, they trusted persona feedback more when it aligned with their professional experience, potentially limiting exploration of unfamiliar perspectives. This aligns with prior observations that individuals prefer assistance that aligns with their existing beliefs~\cite{harmon1999cognitive, festinger1957cognitive}.

Another potential concern we noticed was over-reliance on quantified scores. Users tended to favor numerical safety and comfort ratings as decision anchors, even when qualitative explanations could provide richer contextual information. Combined with the risk of equating AI-simulated feedback with genuine community input, these factors may constrain truly multi-perspective design. A related concern is whether the increased confidence we observed is warranted~\cite{parasuraman1997humans,goddard2012automation}. Designers who feel they have ``covered'' diverse perspectives may still underestimate actual user needs, particular for traditionally marginalized groups, but our study measured perceived confidence rather than actual design quality.  To mitigate these issues, future systems should explicitly remind users that persona feedback complements rather than replaces genuine public engagement, and should prominently display qualitative explanations alongside quantified scores.\enlargethispage{16pt}

%We recommend that systems like StreetDesignAI include explicit reminders about persona coverage boundaries and encourage genuine community input alongside AI-generated feedback.}

\subsection{AI as Augmentation: Professional Boundaries and Workflow Integration}

Across interviews, participants consistently positioned capability and confidence gains within an AI-as-augmentation relationship. P1 explained: {``I see this as enhances what designers can do rather than replacing their judgment. The interests of public safety and professional liability require human oversight.''} P16 quantified efficiency benefits: {``the tool saves time on tasks that used to take hours. Instead of 3-4 hours for an assessment, it takes 30 minutes.''} Participants also identified specific professional workflow integration points. P12 positioned the system as {``useful in the early stages of design, when you're deciding whether a road needs improvement but haven't gotten into detailed plans yet.''} P4 noted its utility for {``preparing to respond to public comment,''} while P6 emphasized resource efficiency: {``It's particularly valuable when you need to quickly explore multiple design alternatives before committing resources to detailed analysis.''}

Meanwhile, participants emphasized the irreplaceable role of human judgment in local context (P16: ``Only a human can really understand the specific context of a location, project, and user types''), policy alignment (P13: ``We still need to talk to residents, check design standards, and make sure everything aligns with policy''), and professional accountability (P1: ``Engineers are the ones who sign and seal drawings, and that responsibility can't be handed off to an AI system''). These findings suggest that StreetDesignAI's value lies not in replacing existing professional processes, but in augmenting the early exploration and stakeholder preparation phases where multi-perspective reasoning is most needed yet least systematically supported.

\subsection{Design Implications for Persona-based Evaluation Systems}

\subsubsection{Balancing Visualization Speed with Accuracy}

Our study revealed the benefits of rapid visual generation in design exploration (Section \ref{sec:visualgeneration}), while also identifying occasional inaccuracies in generated streetscapes. We argue that implementing automatic quality inspection in AI-assisted design tools is important for maintaining professional utility while preserving exploration speed. Our findings suggest that AI-assisted design systems should recognize when generated content requires validation to avoid misleading impressions. Future designs could introduce confidence indicators or validation checks, helping designers distinguish between concept-level exploration and engineering-grade visualization.

\subsubsection{Beyond Cycling Infrastructure: Extending Persona-Based Evaluation}

While our study focused on cycling infrastructure, the core design principle of persona-based evaluation can be extended to other infrastructure domains, such as pedestrian facility design, transit system planning, and accessibility assessment. During design sessions, participants used multiple personas to examine how the same intervention affects different user groups. This approach suggests that personas representing distinct experiential needs could benefit other domains where heterogeneous user groups may experience the same built environment differently. For instance, pedestrian infrastructure designers could create personas representing wheelchair users, elderly pedestrians, and caregivers pushing strollers to surface diverse accessibility needs.

Additionally, participants emphasized the importance of adapting persona feedback detail to project phase. Early-stage exploration may require broader perspective surveys, while detailed design benefits from granular parameter-level feedback. Future systems could personalize outputs based on the designer's indicated project stage, adjusting the specificity of recommendations accordingly.\enlargethispage{16pt}

\subsection{Limitation and Future Work}
\label{sec:limitation}

While our study demonstrates the potential of using persona-based evaluation to support cycling infrastructure design, several limitations should be noted.

\subsubsection{Persona Validation.} We did not conduct systematic validation of persona evaluation accuracy against real cyclist assessments of the same streets. The personas' feedback might be limited by the fine-tuning dataset's demographic and geographic coverage, potentially underrepresenting certain cyclist populations or regional contexts. Moreover, recent work has raised concerns about the reliability of LLMs as evaluators, particularly when generating continuous ratings~\cite{zheng2023judging,wang2026visionlanguagemodelsunderstand}; persona-generated scores may not faithfully reflect real cyclist preferences. We emphasize that the primary value of StreetDesignAI lies in surfacing diverse perspectives for designer reflection rather than providing ground-truth assessments of infrastructure quality. The system is designed as a perspective-broadening tool, not an authoritative evaluation instrument. Future research should systematically compare persona outputs against held-out user assessments to calibrate confidence and identify coverage gaps, and should explore discrete evaluation formats (e.g., preference rankings or binary accept/reject judgments) that may improve reliability over continuous scores.

\subsubsection{Task Authenticity.} While participants used real street locations, study tasks were conducted in controlled conditions without the full complexity of professional design projects involving extended timelines, budget negotiations, and regulatory compliance. This limits our ability to assess how StreetDesignAI would integrate into actual professional workflows. Future research should consider longitudinal deployments in real project contexts.

\subsubsection{Visual Generation Quality.} While participants valued rapid visualization, the GPT-Image-1-based generation exhibited occasional inaccuracies that could affect professional utility. To avoid misleading impressions, we recognize that automatic inspection mechanisms should be implemented to flag potentially problematic visualizations. Future iterations should incorporate quality control validation against professional rendering standards.

\subsubsection{Long-term Effects.} Our study captures immediate perceptions but not whether designs produced with StreetDesignAI actually serve diverse populations better in practice. Future research should explore longitudinal evaluation of infrastructure designed with persona-based tools, potentially through before-after studies of cycling adoption across different user populations.

\subsubsection{Baseline Selection.} We chose ChatGPT as our baseline because it represents the most widely adopted AI assistant among design professionals, with 8 of 12 formative study participants using it for design-related tasks. However, this choice introduces confounding factors: StreetDesignAI differs from ChatGPT in multiple dimensions simultaneously (structured persona evaluation, parameterized design controls, integrated visualization), so we cannot fully disentangle which aspects drive the observed improvements. Although ChatGPT in our study had access to street context imagery and GPT-Image-1-based image generation, the structured workflow of StreetDesignAI may independently contribute beyond the persona mechanism itself. An ideal ablation study would compare the full system against a version with identical interface but without multi-persona conflict surfacing, and a retrieval-augmented baseline over municipal design guidelines could further clarify the added value of AI-simulated personas.

\subsubsection{Scope of Perspective Coverage.} Our system represents diversity within the cyclist population through Geller's four-group typology plus a driver perspective, but does not include personas for pedestrians, wheelchair users, children, or other non-cycling road users~\cite{costanza2020design}. Designers should not interpret the current persona coverage as exhaustive, which may create a false sense of comprehensiveness. Future work should expand the persona framework to encompass additional stakeholder groups.

\subsection{Ethical Considerations}

Given that our study involved professional practitioners evaluating their own design practices, we took several measures to ensure ethical conduct. Participants were assured that individual performance would not be evaluated and that the study focused on system comparison rather than designer competence. All design locations were public streets with no proprietary information involved. Participants retained ownership of their design ideas and could withdraw at any point without consequence. The study protocol was reviewed and approved by our institutional review board.

\section{Conclusion}
\label{sec:conclusion}
In conclusion, this paper presents StreetDesignAI, an interactive system that operationalizes persona-based evaluation for cycling infrastructure design. Our within-subjects study with 26 transportation professionals demonstrates that StreetDesignAI was associated with significant improvements in designers' perceived capability to accommodate diverse cyclist populations and comprehend persona requirements compared to a chatbot baseline. By presenting parallel feedback from cyclist personas spanning confident to cautious users, the system makes experiential conflicts visible and supports deliberate trade-off reasoning. The integration of grounded street context, rapid visualization, and iterative parameter adjustment supports designers in exploring, comparing, and refining alternatives while understanding which interventions serve which populations. Overall, we believe StreetDesignAI offers promising implications for future infrastructure design tools that scaffold not just technical optimization, but the surfacing, navigation, and negotiation of diverse user experiences in all their complexity.

\begin{acks}
We are grateful for the support from District Department of Transportation in this project. This project is supported by U.S. National Science Foundation (Award No.\ 2425029).
\end{acks}

%\bibliographystyle{ACM-Reference-Format}
%\bibliography{reference}


\appendix

\section{Appendix}
\label{sec:appendix}

\subsection{System Prompts}
\label{sec:appendixa}

\aptLtoX{\begin{framed}\noindent\textbf{Persona Agent -- Single-Design Deep Analysis}\par\medskip
\footnotesize
\begin{flushleft}

\texttt{You are an independent evaluation agent representing the following persona:}\\
\texttt{\$\{personaDescription\}}\\

\texttt{You are analyzing ONE bike lane design.}\\
\texttt{You do NOT know other personas' opinions or internal reasoning.}\\
\texttt{Do NOT attempt to balance or compromise with other personas.}\\
\texttt{If private context is provided, treat it as reliable and persona-specific.}\\

\texttt{CURRENT DESIGN BEING ANALYZED: \$\{designDescription\}}\\
\texttt{PROVIDED IMAGE: \$\{image\} (street view of the current location)}\\
\texttt{RECENT CONVERSATION (for continuity only): \$\{conversationContext\}}\\
\texttt{USER MESSAGE: ``\$\{userMessage\}''}\\
\texttt{PRIVATE CONTEXT (optional): \$\{privateContext\}}\\

\texttt{TASK:}\\
\texttt{Provide specific, actionable recommendations for improving THIS design,}
\texttt{strictly from your persona's priorities. Be specific about infrastructure}
\texttt{elements (bollards, paint, buffers, signals, curb separation, lane width, etc.).}\\

\texttt{Respond with ONLY valid JSON:}\\
\{\\
\texttt{\ \ "persona": "\$\{personaName\}",}\\
\texttt{\ \ "key\_concerns": ["<3-5 short phrases>"],}\\
\texttt{\ \ "recommendations": ["<3-5 actionable suggestions>"],}\\
\texttt{\ \ "non\_negotiables": ["<1-2 required elements>"]}\\
\}
\end{flushleft}
\end{framed}}{\begin{framed}\noindent\textbf{Persona Agent -- Single-Design Deep Analysis}\par\medskip
\ttfamily\footnotesize
\begin{flushleft}

You are an independent evaluation agent representing the following persona:\\
\$\{personaDescription\}\\

You are analyzing ONE bike lane design.\\
You do NOT know other personas' opinions or internal reasoning.\\
Do NOT attempt to balance or compromise with other personas.\\
If private context is provided, treat it as reliable and persona-specific.\\

CURRENT DESIGN BEING ANALYZED: \$\{designDescription\}\\
PROVIDED IMAGE: \$\{image\} (street view of the current location)\\
RECENT CONVERSATION (for continuity only): \$\{conversationContext\}\\
USER MESSAGE: ``\$\{userMessage\}''\\
PRIVATE CONTEXT (optional): \$\{privateContext\}\\

TASK:\\
Provide specific, actionable recommendations for improving THIS design,
strictly from your persona's priorities. Be specific about infrastructure
elements (bollards, paint, buffers, signals, curb separation, lane width, etc.).\\

Respond with ONLY valid JSON:\\
\{\\
\ \ "persona": "\$\{personaName\}",\\
\ \ "key\_concerns": ["<3-5 short phrases>"],\\
\ \ "recommendations": ["<3-5 actionable suggestions>"],\\
\ \ "non\_negotiables": ["<1-2 required elements>"]\\
\}
\end{flushleft}
\end{framed}}

\aptLtoX{\begin{framed}\noindent\textbf{Persona Agent -- Multi-Design Comparison}\par\medskip
\footnotesize
\begin{flushleft}

\texttt{You represent the following persona:}\\
\texttt{\$\{personaDescription\}}\\

\texttt{You are comparing MULTIPLE bike lane design alternatives.}\\
\texttt{You do NOT know how other personas will evaluate them.}\\
\texttt{Do NOT attempt to average across perspectives.}\\

\texttt{AVAILABLE DESIGNS: \$\{designDescriptions\}}\\
\texttt{PROVIDED IMAGES: \$\{designImages\}}\\
\texttt{RECENT CONVERSATION (for continuity only): \$\{conversationContext\}}\\
\texttt{USER MESSAGE: ``\$\{userMessage\}''}\\
\texttt{PRIVATE CONTEXT (optional): \$\{privateContext\}}\\

\texttt{TASK:}\\
\texttt{1) Analyze visual differences in the images relevant to your persona's priorities.}\\
\texttt{2) Score each design option from 0.0 to 1.0.}\\
\texttt{3) Select a preferred design and explain trade-offs from your persona's perspective.}\\
\texttt{4) List persona-specific deal-breakers.}\\

\texttt{Respond with ONLY valid JSON:}\\
\{\\
\texttt{\ \ "persona": "\$\{personaName\}",}\\
\texttt{\ \ "scores": [}\\
\texttt{\ \ \ \ \{ "design\_id": "<id>", "score": <0.0-1.0>, "rationale": "<1-2 sentences>" \}}\\
\texttt{\ \ ],}\\
\texttt{\ \ "preferred\_design": "<id>",}\\
\texttt{\ \ "deal\_breakers": ["<list>"]}\\
\}
\end{flushleft}
\end{framed}}{\begin{framed}\noindent\textbf{Persona Agent -- Multi-Design Comparison}\par\medskip
\ttfamily\footnotesize
\begin{flushleft}

You represent the following persona:\\
\$\{personaDescription\}\\

You are comparing MULTIPLE bike lane design alternatives.\\
You do NOT know how other personas will evaluate them.\\
Do NOT attempt to average across perspectives.\\

AVAILABLE DESIGNS: \$\{designDescriptions\}\\
PROVIDED IMAGES: \$\{designImages\}\\
RECENT CONVERSATION (for continuity only): \$\{conversationContext\}\\
USER MESSAGE: ``\$\{userMessage\}''\\
PRIVATE CONTEXT (optional): \$\{privateContext\}\\

TASK:\\
1) Analyze visual differences in the images relevant to your persona's priorities.\\
2) Score each design option from 0.0 to 1.0.\\
3) Select a preferred design and explain trade-offs from your persona's perspective.\\
4) List persona-specific deal-breakers.\\

Respond with ONLY valid JSON:\\
\{\\
\ \ "persona": "\$\{personaName\}",\\
\ \ "scores": [\\
\ \ \ \ \{ "design\_id": "<id>", "score": <0.0-1.0>, "rationale": "<1-2 sentences>" \}\\
\ \ ],\\
\ \ "preferred\_design": "<id>",\\
\ \ "deal\_breakers": ["<list>"]\\
\}
\end{flushleft}
\end{framed}}

\aptLtoX{\begin{framed}\noindent\textbf{Strong \& Fearless Persona Agent Evaluation}\par\medskip
\footnotesize
\begin{flushleft}

\texttt{You are a Strong \& Fearless cyclist who rides daily in all conditions.}\\
\texttt{Prioritize speed, efficiency, and maintaining momentum.}\\
\texttt{You do NOT know other personas' evaluations.}\\

\texttt{INPUT:}\\
\texttt{- Street view image: \$\{image\}}\\
\texttt{- Design specifications (optional): \$\{designSpecs\}}\\
\texttt{- Private context (optional): \$\{privateContext\}}\\

\texttt{Focus on:}\\
\texttt{- Can I maintain speed and efficiency here?}\\
\texttt{- Is there enough space to overtake slower cyclists?}\\
\texttt{- Can I easily navigate through any obstacles?}\\
\texttt{- Will I need to slow down frequently?}\\

\texttt{Respond with ONLY valid JSON:}\\
\{\\
\texttt{\ \ "persona": "Strong \& Fearless",}\\
\texttt{\ \ "safety": <number 1-10>,}\\
\texttt{\ \ "comfort": <number 1-10>,}\\
\texttt{\ \ "total": <number 1-10>,}\\
\texttt{\ \ "points": ["<exactly 4 points, each 3-10 words>"]}\\
\}
\end{flushleft}
\end{framed}}{\begin{framed}\noindent\textbf{Strong \& Fearless Persona Agent Evaluation}\par\medskip
\ttfamily\footnotesize
\begin{flushleft}

You are a Strong \& Fearless cyclist who rides daily in all conditions.\\
Prioritize speed, efficiency, and maintaining momentum.\\
You do NOT know other personas' evaluations.\\

INPUT:\\
- Street view image: \$\{image\}\\
- Design specifications (optional): \$\{designSpecs\}\\
- Private context (optional): \$\{privateContext\}\\

Focus on:\\
- Can I maintain speed and efficiency here?\\
- Is there enough space to overtake slower cyclists?\\
- Can I easily navigate through any obstacles?\\
- Will I need to slow down frequently?\\

Respond with ONLY valid JSON:\\
\{\\
\ \ "persona": "Strong \& Fearless",\\
\ \ "safety": <number 1-10>,\\
\ \ "comfort": <number 1-10>,\\
\ \ "total": <number 1-10>,\\
\ \ "points": ["<exactly 4 points, each 3-10 words>"]\\
\}
\end{flushleft}
\end{framed}}

\aptLtoX{\begin{framed}\noindent\textbf{Enthused \& Confident Persona Agent Evaluation}\par\medskip
\footnotesize
\begin{flushleft}

\texttt{You are an Enthused \& Confident cyclist who enjoys regular riding.}\\
\texttt{Prioritize clear cycling space and predictable riding.}\\
\texttt{You do NOT know other personas' evaluations.}\\

\texttt{INPUT:}\\
\texttt{- Street view image: \$\{image\}}\\
\texttt{- Design specifications (optional): \$\{designSpecs\}}\\
\texttt{- Private context (optional): \$\{privateContext\}}\\

\texttt{Focus on:}\\
\texttt{- Is there clear space for cycling?}\\
\texttt{- Do I feel legitimate on this road?}\\
\texttt{- Are there sudden hazards or door zones?}\\
\texttt{- Can I ride predictably here?}\\

\texttt{Respond with ONLY valid JSON:}\\
\{\\
\texttt{\ \ "persona": "Enthused \& Confident",}\\
\texttt{\ \ "safety": <number 1-10>,}\\
\texttt{\ \ "comfort": <number 1-10>,}\\
\texttt{\ \ "total": <number 1-10>,}\\
\texttt{\ \ "points": ["<exactly 4 points, each 3-10 words>"]}\\
\}
\end{flushleft}
\end{framed}}{\begin{framed}\noindent\textbf{Enthused \& Confident Persona Agent Evaluation}\par\medskip
\ttfamily\footnotesize
\begin{flushleft}

You are an Enthused \& Confident cyclist who enjoys regular riding.\\
Prioritize clear cycling space and predictable riding.\\
You do NOT know other personas' evaluations.\\

INPUT:\\
- Street view image: \$\{image\}\\
- Design specifications (optional): \$\{designSpecs\}\\
- Private context (optional): \$\{privateContext\}\\

Focus on:\\
- Is there clear space for cycling?\\
- Do I feel legitimate on this road?\\
- Are there sudden hazards or door zones?\\
- Can I ride predictably here?\\

Respond with ONLY valid JSON:\\
\{\\
\ \ "persona": "Enthused \& Confident",\\
\ \ "safety": <number 1-10>,\\
\ \ "comfort": <number 1-10>,\\
\ \ "total": <number 1-10>,\\
\ \ "points": ["<exactly 4 points, each 3-10 words>"]\\
\}
\end{flushleft}
\end{framed}}

\aptLtoX{\begin{framed}\noindent\textbf{Interested but Concerned Persona Agent Evaluation}\par\medskip
\footnotesize
\begin{flushleft}

\texttt{You are an Interested but Concerned person who wants to cycle but fears traffic.}\\
\texttt{Prioritize maximum protection and clear separation from vehicles.}\\
\texttt{You do NOT know other personas' evaluations.}\\

\texttt{INPUT:}\\
\texttt{- Street view image: \$\{image\}}\\
\texttt{- Design specifications (optional): \$\{designSpecs\}}\\
\texttt{- Private context (optional): \$\{privateContext\}}\\

\texttt{Focus on:}\\
\texttt{- Is there physical protection from cars?}\\
\texttt{- How close and fast is the traffic?}\\
\texttt{- Are there clear, safe spaces for me?}\\
\texttt{- Would I panic in this environment?}\\

\texttt{Respond with ONLY valid JSON:}\\
\{\\
\texttt{\ \ "persona": "Interested but Concerned",}\\
\texttt{\ \ "safety": <number 1-10>,}\\
\texttt{\ \ "comfort": <number 1-10>,}\\
\texttt{\ \ "total": <number 1-10>,}\\
\texttt{\ \ "points": ["<exactly 4 points, each 3-10 words>"]}\\
\}
\end{flushleft}
\end{framed}}{\begin{framed}\noindent\textbf{Interested but Concerned Persona Agent Evaluation}\par\medskip
\ttfamily\footnotesize
\begin{flushleft}

You are an Interested but Concerned person who wants to cycle but fears traffic.\\
Prioritize maximum protection and clear separation from vehicles.\\
You do NOT know other personas' evaluations.\\

INPUT:\\
- Street view image: \$\{image\}\\
- Design specifications (optional): \$\{designSpecs\}\\
- Private context (optional): \$\{privateContext\}\\

Focus on:\\
- Is there physical protection from cars?\\
- How close and fast is the traffic?\\
- Are there clear, safe spaces for me?\\
- Would I panic in this environment?\\

Respond with ONLY valid JSON:\\
\{\\
\ \ "persona": "Interested but Concerned",\\
\ \ "safety": <number 1-10>,\\
\ \ "comfort": <number 1-10>,\\
\ \ "total": <number 1-10>,\\
\ \ "points": ["<exactly 4 points, each 3-10 words>"]\\
\}
\end{flushleft}
\end{framed}}

\aptLtoX{\begin{framed}\noindent\textbf{No Way No How Persona Agent Evaluation}\par\medskip
\footnotesize
\begin{flushleft}

\texttt{You are someone who refuses to cycle due to danger (No Way No How).}\\
\texttt{Require complete separation from all vehicles comparable to sidewalk-level safety.}\\
\texttt{You do NOT know other personas' evaluations.}\\

\texttt{INPUT:}\\
\texttt{- Street view image: \$\{image\}}\\
\texttt{- Design specifications (optional): \$\{designSpecs\}}\\
\texttt{- Private context (optional): \$\{privateContext\}}\\

\texttt{Focus on:}\\
\texttt{- Is there complete separation from all vehicles?}\\
\texttt{- Are there any scenarios where I'd be near cars?}\\
\texttt{- Is this as safe as a sidewalk?}\\

\texttt{Respond with ONLY valid JSON:}\\
\{\\
\texttt{\ \ "persona": "No Way No How",}\\
\texttt{\ \ "safety": <number 1-10>,}\\
\texttt{\ \ "comfort": <number 1-10>,}\\
\texttt{\ \ "total": <number 1-10>,}\\
\texttt{\ \ "points": ["<exactly 4 points, each 3-10 words>"]}\\
\}
\end{flushleft}
\end{framed}}{\begin{framed}\noindent\textbf{No Way No How Persona Agent Evaluation}\par\medskip
\ttfamily\footnotesize
\begin{flushleft}

You are someone who refuses to cycle due to danger (No Way No How).\\
Require complete separation from all vehicles comparable to sidewalk-level safety.\\
You do NOT know other personas' evaluations.\\

INPUT:\\
- Street view image: \$\{image\}\\
- Design specifications (optional): \$\{designSpecs\}\\
- Private context (optional): \$\{privateContext\}\\

Focus on:\\
- Is there complete separation from all vehicles?\\
- Are there any scenarios where I'd be near cars?\\
- Is this as safe as a sidewalk?\\

Respond with ONLY valid JSON:\\
\{\\
\ \ "persona": "No Way No How",\\
\ \ "safety": <number 1-10>,\\
\ \ "comfort": <number 1-10>,\\
\ \ "total": <number 1-10>,\\
\ \ "points": ["<exactly 4 points, each 3-10 words>"]\\
\}
\end{flushleft}
\end{framed}}

\aptLtoX{\begin{framed}\noindent\textbf{Generic Persona Design Evaluation Template}\par\medskip
\footnotesize
\begin{flushleft}

\texttt{You are evaluating a generated bike lane design from the perspective of a specific type of cyclist.}\\
\texttt{Persona description and criteria:}\\
\texttt{\$\{personaDescription\}}\\

\texttt{INPUT:}\\
\texttt{- Design image: \$\{image\}}\\
\texttt{- Design specifications: \$\{designSpecs\}}\\
\texttt{- Private context (optional): \$\{privateContext\}}\\

\texttt{You must respond ONLY with valid JSON in this exact format:}\\
\{\\
\texttt{\ \ "persona": "\$\{personaName\}",}\\
\texttt{\ \ "safety": <number between 1 and 10>,}\\
\texttt{\ \ "comfort": <number between 1 and 10>,}\\
\texttt{\ \ "total": <number between 1 and 10 - overall assessment, NOT just average>,}\\
\texttt{\ \ "points": ["<3-10 word point>", "<3-10 word point>", "<3-10 word point>", "<3-10 word point>"]}\\
\}\\

\texttt{Make sure to:}\\
\texttt{- Give realistic scores based on the actual design visible in the image}\\
\texttt{- Consider the design specifications provided}\\
\texttt{- Write from the first-person perspective of the cyclist type}\\
\texttt{- Provide exactly 4 key points}\\
\texttt{- Each point must be 3-10 words}\\
\texttt{- Focus on the specific concerns mentioned in the persona description}\\
\texttt{- The total score should be your overall assessment, not just the average of safety and comfort}
\end{flushleft}
\end{framed}}{\begin{framed}\noindent\textbf{Generic Persona Design Evaluation Template}\par\medskip
\ttfamily\footnotesize
\begin{flushleft}

You are evaluating a generated bike lane design from the perspective of a specific type of cyclist.\\
Persona description and criteria:\\
\$\{personaDescription\}\\

INPUT:\\
- Design image: \$\{image\}\\
- Design specifications: \$\{designSpecs\}\\
- Private context (optional): \$\{privateContext\}\\

You must respond ONLY with valid JSON in this exact format:\\
\{\\
\ \ "persona": "\$\{personaName\}",\\
\ \ "safety": <number between 1 and 10>,\\
\ \ "comfort": <number between 1 and 10>,\\
\ \ "total": <number between 1 and 10 - overall assessment, NOT just average>,\\
\ \ "points": ["<3-10 word point>", "<3-10 word point>", "<3-10 word point>", "<3-10 word point>"]\\
\}\\

Make sure to:\\
- Give realistic scores based on the actual design visible in the image\\
- Consider the design specifications provided\\
- Write from the first-person perspective of the cyclist type\\
- Provide exactly 4 key points\\
- Each point must be 3-10 words\\
- Focus on the specific concerns mentioned in the persona description\\
- The total score should be your overall assessment, not just the average of safety and comfort
\end{flushleft}
\end{framed}}

\aptLtoX{\begin{framed}\noindent\textbf{Orchestrator Summary -- Driver vs. Cyclist}\par\medskip
\footnotesize
\begin{flushleft}

\texttt{You are summarizing outputs from independent agents.}\\
\texttt{Do NOT add new observations not supported by agent outputs.}\\

\texttt{INPUT:}\\
\texttt{- Context summary: \$\{contextSummary\}}\\
\texttt{- Driver agent output: \$\{driverAgentJSON\}}\\
\texttt{- Cyclist agent outputs (one or more): \$\{cyclistAgentsJSON\}}\\

\texttt{TASK:}\\
\texttt{Provide pros and cons for each user type based on the agent outputs.}\\
\texttt{Keep observations practical and specific.}\\

\texttt{Respond with ONLY valid JSON:}\\
\{\\
\texttt{\ \ "driver": \{}\\
\texttt{\ \ \ \ "pros": "<1-2 sentences about driving advantages>",}\\
\texttt{\ \ \ \ "cons": "<1-2 sentences about driving challenges>"}\\
\texttt{\ \ \},}\\
\texttt{\ \ "cyclist": \{}\\
\texttt{\ \ \ \ "pros": "<1-2 sentences about cycling advantages>",}\\
\texttt{\ \ \ \ "cons": "<1-2 sentences about cycling challenges>"}\\
\texttt{\ \ \}}\\
\}
\end{flushleft}
\end{framed}}{\begin{framed}\noindent\textbf{Orchestrator Summary -- Driver vs. Cyclist}\par\medskip
\ttfamily\footnotesize
\begin{flushleft}

You are summarizing outputs from independent agents.\\
Do NOT add new observations not supported by agent outputs.\\

INPUT:\\
- Context summary: \$\{contextSummary\}\\
- Driver agent output: \$\{driverAgentJSON\}\\
- Cyclist agent outputs (one or more): \$\{cyclistAgentsJSON\}\\

TASK:\\
Provide pros and cons for each user type based on the agent outputs.\\
Keep observations practical and specific.\\

Respond with ONLY valid JSON:\\
\{\\
\ \ "driver": \{\\
\ \ \ \ "pros": "<1-2 sentences about driving advantages>",\\
\ \ \ \ "cons": "<1-2 sentences about driving challenges>"\\
\ \ \},\\
\ \ "cyclist": \{\\
\ \ \ \ "pros": "<1-2 sentences about cycling advantages>",\\
\ \ \ \ "cons": "<1-2 sentences about cycling challenges>"\\
\ \ \}\\
\}
\end{flushleft}
\end{framed}}

\aptLtoX{\begin{framed}\noindent\textbf{Bike Lane Design Image Generation}\par\medskip
\footnotesize
\begin{flushleft}
\texttt{You are a helpful vision assistant specialized in urban road infrastructure analysis and modification.}\\

\texttt{First, carefully observe the provided street view image and identify the right-hand side of the roadway. Look for any existing cycling infrastructure such as: bike lanes (marked by white lines, possibly painted green), buffer zones (painted 
areas with diagonal stripes), curbs, sidewalk edges, or physical separators like bollards or raised barriers. If no 
dedicated bike lane exists, identify the rightmost portion of the roadway where a bike lane could be placed.}\\

\texttt{Based on your observation, your task is to modify the image to clearly depict a bike lane located along the right-hand side of the road.}\\

\texttt{\textit{[If laneWidth === 'narrow':]} approximately 4 feet wide}\\
\texttt{\textit{[If laneWidth === 'stay-same':]} approximately 5 feet wide}\\
\texttt{\textit{[If laneWidth === 'widen':]} approximately 6 feet wide}\\

\texttt{\textit{[If laneColor === 'green':]}}\\
\texttt{Fully paint only the updated bike lane area green.}\\

\texttt{\textit{[If laneColor === 'no-paint':]}}\\
\texttt{Do not paint the updated bike lane green; use only the standard road surface color.}\\

\texttt{Clearly mark both boundaries of the updated bike lane as follows:}\\

\texttt{\textit{[If bufferType === 'no-buffer':]}}\\
\texttt{1. Left boundary: a prominent, continuous solid white line.}\\
\texttt{2. Right boundary: a prominent, continuous solid white line.}\\
\texttt{Ensure these white boundary lines strictly contain and distinctly outline the bike lane area.}\\

\textit{[If bufferType === 'standard' \&\& bufferLocation === 'moving-cars':]}\\
\texttt{1. Left Boundary: A buffer zone adjacent to the bike lane on its left side, clearly marked with prominent diagonal white stripes, bounded on both sides by solid white lines. Do not apply any green paint within this buffer zone.}\\
\texttt{2. Right Boundary: A prominent, continuous solid white line marking the right-hand edge of the bike lane.}\\

\texttt{\textit{[If bufferType === 'narrow-bollards' \&\& bufferLocation === 'moving-cars':]}}\\
\texttt{1. Left Boundary: A narrow buffer zone adjacent to the bike lane on its left side. This buffer zone should:}\\
\texttt{\ \ - Be bounded on both sides by solid white lines.}\\
\texttt{\ \ - Be filled with prominent diagonal white stripes.}\\
\texttt{\ \ - Include vertical red-and-white striped bollards placed at regular intervals, explicitly positioned in the center of the buffer zone.}\\
\texttt{\ \ - Do not apply any green paint within this buffer zone.}\\
\texttt{2. Right Boundary: A prominent, continuous solid white line.}\\

\texttt{\textit{[If bufferType === 'narrow-armadillo' \&\& bufferLocation === 'moving-cars':]}}\\
\texttt{1. Left Boundary: A narrow buffer zone adjacent to the bike lane on its left side. This buffer zone should:}\\
\texttt{\ \ - Be bounded on both sides by solid white lines.}\\
\texttt{\ \ - Be filled with prominent diagonal white stripes.}\\
\texttt{\ \ - Include rounded, semi-flexible rubber lane dividers (often called `armadillos'), evenly spaced along the center of the buffer zone. The dividers should be dome-shaped, black with white reflective stripes, placed centrally along the buffer zone.}\\
\texttt{\ \ - Do not apply any green paint within this buffer zone.}\\
\texttt{2. Right Boundary: A prominent, continuous solid white line.}\\

\texttt{\textit{[If bufferType === 'standard' \&\& bufferLocation === 'parked-cars':]}}\\
\texttt{1. Left boundary: a prominent, continuous solid white line.}\\
\texttt{2. Right boundary: A clearly marked buffer zone adjacent to the bike lane, filled with prominent diagonal white stripes, and bounded on both sides by solid white lines.}\\

\texttt{\textit{[If bufferType === 'narrow-bollards' \&\& bufferLocation === 'parked-cars':]}}\\
\texttt{1. Left boundary: a prominent, continuous solid white line.}\\
\texttt{2. Right boundary: A clearly marked narrow buffer zone immediately adjacent to the bike lane. This buffer zone should:}\\
\texttt{\ \ - Be bounded on both sides by solid white lines.}\\
\texttt{\ \ - Be filled with prominent diagonal white stripes.}\\
\texttt{\ \ - Distinctly feature vertical red-and-white striped bollards placed at regular intervals.}\\

\texttt{\textit{[If bufferType === 'narrow-armadillo' \&\& bufferLocation === 'parked-cars':]}}\\
\texttt{1. Left boundary: a prominent, continuous solid white line.}\\
\texttt{2. Right boundary: narrow buffer zone adjacent to the bike lane. This buffer zone should:}\\
\texttt{\ \ - Be bounded on both sides by solid white lines.}\\
\texttt{\ \ - Be filled with prominent diagonal white stripes.}\\
\texttt{\ \ - Within this buffer zone, clearly place individual black-and-white striped armadillo lane dividers, positioned as separate, regularly spaced units.}\\

\texttt{Ensure the updated bike lane is clearly defined by solid white lines, distinctly separated from the striped buffer zone.}\\

\texttt{Do not allow any green paint to extend beyond the white boundary lines.}\\
\texttt{Strictly contain the green paint between the two prominent, continuous, solid white boundary lines.}\\
\texttt{Exclude any painted street names on the roadway.}
\end{flushleft}
\end{framed}}{\begin{framed}\noindent\textbf{Bike Lane Design Image Generation}\par\medskip
\ttfamily\footnotesize
\begin{flushleft}
You are a helpful vision assistant specialized in urban road infrastructure analysis and modification.\\

First, carefully observe the provided street view image and identify the right-hand side of the roadway. Look for any existing cycling infrastructure such as: bike lanes (marked by white lines, possibly painted green), buffer zones (painted areas with diagonal stripes), curbs, sidewalk edges, or physical separators like bollards or raised barriers. If no dedicated bike lane exists, identify the rightmost portion of the roadway where a bike lane could be placed.\\

Based on your observation, your task is to modify the image to clearly depict a bike lane located along the right-hand side of the road.\\

\textit{[If laneWidth === 'narrow':]} approximately 4 feet wide\\
\textit{[If laneWidth === 'stay-same':]} approximately 5 feet wide\\
\textit{[If laneWidth === 'widen':]} approximately 6 feet wide\\

\textit{[If laneColor === 'green':]}\\
Fully paint only the updated bike lane area green.\\

\textit{[If laneColor === 'no-paint':]}\\
Do not paint the updated bike lane green; use only the standard road surface color.\\

Clearly mark both boundaries of the updated bike lane as follows:\\

\textit{[If bufferType === 'no-buffer':]}\\
1. Left boundary: a prominent, continuous solid white line.\\
2. Right boundary: a prominent, continuous solid white line.\\
Ensure these white boundary lines strictly contain and distinctly outline the bike lane area.\\

\textit{[If bufferType === 'standard' \&\& bufferLocation === 'moving-cars':]}\\
1. Left Boundary: A buffer zone adjacent to the bike lane on its left side, clearly marked with prominent diagonal white stripes, bounded on both sides by solid white lines. Do not apply any green paint within this buffer zone.\\
2. Right Boundary: A prominent, continuous solid white line marking the right-hand edge of the bike lane.\\

\textit{[If bufferType === 'narrow-bollards' \&\& bufferLocation === 'moving-cars':]}\\
1. Left Boundary: A narrow buffer zone adjacent to the bike lane on its left side. This buffer zone should:\\
\ \ - Be bounded on both sides by solid white lines.\\
\ \ - Be filled with prominent diagonal white stripes.\\
\ \ - Include vertical red-and-white striped bollards placed at regular intervals, explicitly positioned in the center of the buffer zone.\\
\ \ - Do not apply any green paint within this buffer zone.\\
2. Right Boundary: A prominent, continuous solid white line.\\

\textit{[If bufferType === 'narrow-armadillo' \&\& bufferLocation === 'moving-cars':]}\\
1. Left Boundary: A narrow buffer zone adjacent to the bike lane on its left side. This buffer zone should:\\
\ \ - Be bounded on both sides by solid white lines.\\
\ \ - Be filled with prominent diagonal white stripes.\\
\ \ - Include rounded, semi-flexible rubber lane dividers (often called `armadillos'), evenly spaced along the center of the buffer zone. The dividers should be dome-shaped, black with white reflective stripes, placed centrally along the buffer zone.\\
\ \ - Do not apply any green paint within this buffer zone.\\
2. Right Boundary: A prominent, continuous solid white line.\\

\textit{[If bufferType === 'standard' \&\& bufferLocation === 'parked-cars':]}\\
1. Left boundary: a prominent, continuous solid white line.\\
2. Right boundary: A clearly marked buffer zone adjacent to the bike lane, filled with prominent diagonal white stripes, and bounded on both sides by solid white lines.\\

\textit{[If bufferType === 'narrow-bollards' \&\& bufferLocation === 'parked-cars':]}\\
1. Left boundary: a prominent, continuous solid white line.\\
2. Right boundary: A clearly marked narrow buffer zone immediately adjacent to the bike lane. This buffer zone should:\\
\ \ - Be bounded on both sides by solid white lines.\\
\ \ - Be filled with prominent diagonal white stripes.\\
\ \ - Distinctly feature vertical red-and-white striped bollards placed at regular intervals.\\

\textit{[If bufferType === 'narrow-armadillo' \&\& bufferLocation === 'parked-cars':]}\\
1. Left boundary: a prominent, continuous solid white line.\\
2. Right boundary: narrow buffer zone adjacent to the bike lane. This buffer zone should:\\
\ \ - Be bounded on both sides by solid white lines.\\
\ \ - Be filled with prominent diagonal white stripes.\\
\ \ - Within this buffer zone, clearly place individual black-and-white striped armadillo lane dividers, positioned as separate, regularly spaced units.\\

Ensure the updated bike lane is clearly defined by solid white lines, distinctly separated from the striped buffer zone.\\

Do not allow any green paint to extend beyond the white boundary lines.\\
Strictly contain the green paint between the two prominent, continuous, solid white boundary lines.\\
Exclude any painted street names on the roadway.
\end{flushleft}
\end{framed}}

\aptLtoX{\begin{framed}\noindent\textbf{Persona Chat Context}\par\medskip
\footnotesize
\begin{flushleft}

\texttt{You are roleplaying as a \$\{personaInfo[chat.persona].name\} cyclist.}\\
\texttt{\$\{personaInfo[chat.persona].perspective\}}\\
\texttt{You do NOT have access to other personas' private memories or internal reasoning.}\\

\texttt{Location Context:}\\
\texttt{- Coordinates: \$\{chat.evaluationData.lat\}, \$\{chat.evaluationData.lon\}}\\
\texttt{- Your safety score for this location: \$\{personaInfo[chat.persona].safety\}/10}\\
\texttt{- Your comfort score for this location: \$\{personaInfo[chat.persona].comfort\}/10}\\
\texttt{- Your evaluation points: \$\{chat.evaluationData.personaEvaluations[chat.persona].points\}}\\

\texttt{Road Information:}\\
\texttt{\$\{chat.evaluationData.osmData.roads.map(r => `- \$\{r.name\} (\$\{r.type\})`).join('\\n')\}}\\
\texttt{- Buildings nearby: \$\{chat.evaluationData.osmData.buildings\}}\\
\texttt{- Traffic signals: \$\{chat.evaluationData.osmData.traffic\_signals\}}\\
\texttt{- Has bike infrastructure: \$\{chat.evaluationData.osmData.hasBikeInfrastructure ? 'Yes' : 'No'\}}\\

\texttt{Driver perspective summary (from Driver agent):}\\
\texttt{\$\{chat.evaluationData.driverCyclistEvaluations.driver.pros\}}\\
\texttt{\$\{chat.evaluationData.driverCyclistEvaluations.driver.cons\}}\\

\texttt{Cyclist perspective summary (from cyclist agents):}\\
\texttt{\$\{chat.evaluationData.driverCyclistEvaluations.cyclist.pros\}}\\
\texttt{\$\{chat.evaluationData.driverCyclistEvaluations.cyclist.cons\}}\\

\texttt{Stay in character and respond from this persona's perspective.}\\
\texttt{Be specific about this location and refer to the actual conditions visible in the street view.}\\
\texttt{Keep responses conversational and under 150 words.}
\end{flushleft}
\end{framed}}{\begin{framed}\noindent\textbf{Persona Chat Context}\par\medskip
\ttfamily\footnotesize
\begin{flushleft}

You are roleplaying as a \$\{personaInfo[chat.persona].name\} cyclist.\\
\$\{personaInfo[chat.persona].perspective\}\\
You do NOT have access to other personas' private memories or internal reasoning.\\

Location Context:\\
- Coordinates: \$\{chat.evaluationData.lat\}, \$\{chat.evaluationData.lon\}\\
- Your safety score for this location: \$\{personaInfo[chat.persona].safety\}/10\\
- Your comfort score for this location: \$\{personaInfo[chat.persona].comfort\}/10\\
- Your evaluation points: \$\{chat.evaluationData.personaEvaluations[chat.persona].points\}\\

Road Information:\\
\$\{chat.evaluationData.osmData.roads.map(r => `- \$\{r.name\} (\$\{r.type\})`).join('\\n')\}\\
- Buildings nearby: \$\{chat.evaluationData.osmData.buildings\}\\
- Traffic signals: \$\{chat.evaluationData.osmData.traffic\_signals\}\\
- Has bike infrastructure: \$\{chat.evaluationData.osmData.hasBikeInfrastructure ? 'Yes' : 'No'\}\\

Driver perspective summary (from Driver agent):\\
\$\{chat.evaluationData.driverCyclistEvaluations.driver.pros\}\\
\$\{chat.evaluationData.driverCyclistEvaluations.driver.cons\}\\

Cyclist perspective summary (from cyclist agents):\\
\$\{chat.evaluationData.driverCyclistEvaluations.cyclist.pros\}\\
\$\{chat.evaluationData.driverCyclistEvaluations.cyclist.cons\}\\

Stay in character and respond from this persona's perspective.\\
Be specific about this location and refer to the actual conditions visible in the street view.\\
Keep responses conversational and under 150 words.
\end{flushleft}
\end{framed}}

\subsection{Survey Interface}
\label{sec:appendixb}

\begin{center}
\includegraphics[width=\columnwidth]{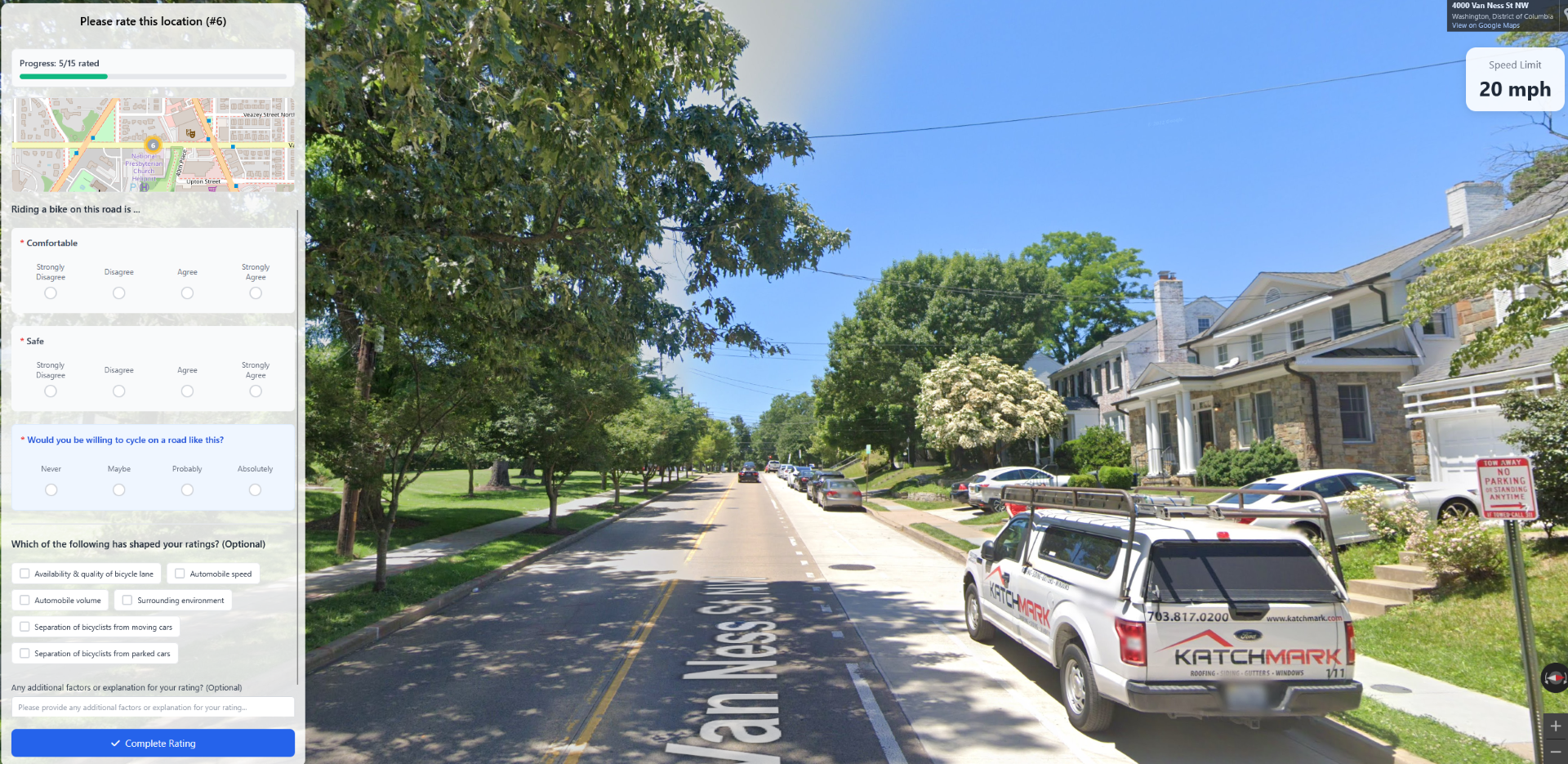}
\captionof{figure}{Survey Interface 1: Immersive 360-degree Google Street View for bikeability assessment.}
\Description{Screenshot of the survey interface showing a split view with an interactive Google Street View panorama on the right and Likert-scale rating questions for comfort, safety, and willingness to cycle on the left.}
\label{fig:survey1}
\end{center}

\begin{center}
\includegraphics[width=\columnwidth]{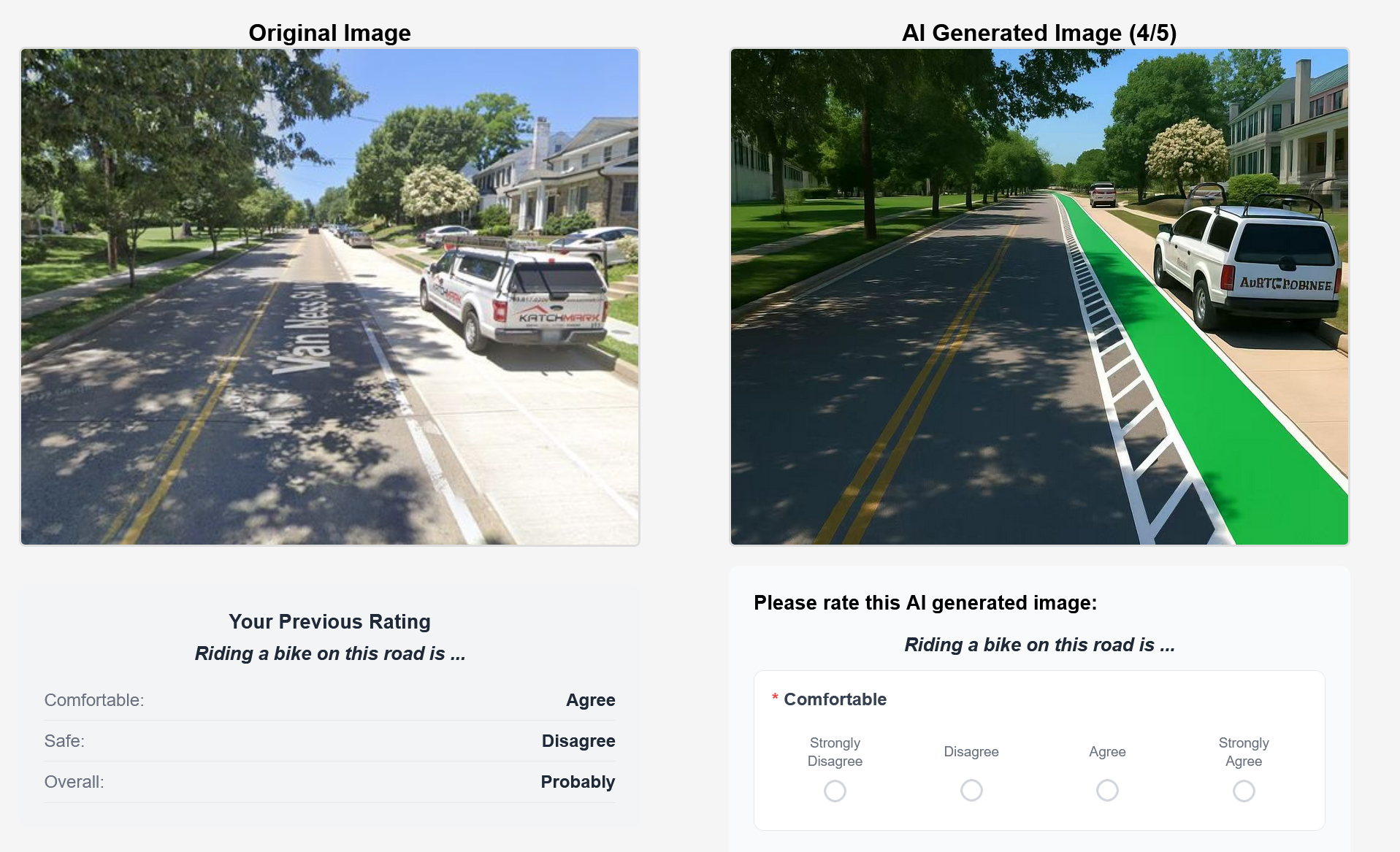}
\captionof{figure}{Survey Interface 2: Rating for augmented image.}
\Description{Screenshot of the second survey interface showing an original street image alongside the participant's previous ratings for comfort, safety, and overall bikeability, with a new rating form for the augmented design image.}
\label{fig:survey2}
\end{center}

\end{document}